\pdfoutput=1
\RequirePackage{ifpdf}
\ifpdf 
\documentclass[pdftex]{sigma}
\else
\documentclass{sigma}
\fi

\IfFileExists{dsfont.sty}
	{\usepackage{dsfont}
 \let\mathbb=\mathds
 \newcommand{\id}{\mathds{1}}}
	{\typeout{Package dsfont.sty was not found, using alternative macros.}
 \let\mathds=\mathbb
 \newcommand{\id}{\mbox{1 \kern-.59em \textrm{l}}}}

\usepackage{mathtools}

\numberwithin{equation}{section}

\newcommand{\nc}{noncommutative}

\newcommand{\starco}[2]{\left[ #1\stackrel{\star}{,}#2\right] }		
\newcommand{\staraco}[2]{\left\{ #1\stackrel{\star}{,}#2\right\} }	
\newcommand{\var}[2]{\frac{\d #1}{\d #2}}				
\newcommand{\pa}{\partial}						

\newcommand{\ri}{\txt{i}}						
\newcommand{\re}{\txt{e}}						

\renewcommand{\d}{\delta}

\renewcommand{\th}{\theta}

\newcommand{\m}{\mu}
\newcommand{\n}{\nu}

\newcommand{\s}{\sigma}

\renewcommand{\Xi}{\Xi}

\newcommand{\ig}{\txt{i}g}

\newcommand{\Tr}{\operatorname{Tr}}

\newcommand{\cO}{\mathcal{O}}

\newcommand{\vp}{\varphi}
\renewcommand{\pa}{\partial}
\renewcommand{\ri}{\textrm{i}}

\newcommand{\txt}[1]{\textrm{#1}}
\newcommand{\br}{\mathbb{R}}

\newcommand{\cg}{c}

\newcommand{\wt}{\texttt{w}}
\newcommand{\cd}{{\cal D}}

\makeatletter
\g@addto@macro\bfseries{\boldmath}
\makeatother

\begin{document}

\allowdisplaybreaks

\newcommand{\arXivNumber}{1806.02131}

\renewcommand{\PaperNumber}{133}

\FirstPageHeading

\ShortArticleName{Field Theory with Coordinate Dependent Noncommutativity}

\ArticleName{Field Theory with Coordinate Dependent\\ Noncommutativity}

\Author{Daniel N. BLASCHKE~$^{\dag^1}$, Fran\c{c}ois GIERES~$^{\dag^2}$, Stefan HOHENEGGER~$^{\dag^2}$, \\ Manfred SCHWEDA~$^{\dag^3}$ and Michael WOHLGENANNT~$^{\dag^4}$}

\AuthorNameForHeading{D.N.~Blaschke, F.~Gieres, S.~Hohenegger, M.~Schweda and M.~Wohlgenannt}

\Address{$^{\dag^1}$~Los Alamos National Laboratory, Los Alamos, NM, 87545, USA}
\EmailDD{\href{mailto:dblaschke@lanl.gov}{dblaschke@lanl.gov}}

\Address{$^{\dag^2}$~Institut de Physique Nucl\'eaire de Lyon, Universit\'e de Lyon,\\
\hphantom{$^{\dag^2}$}~Universit\'e Claude Bernard Lyon~1 and CNRS/IN2P3, Bat. Paul Dirac, 4 rue Enrico Fermi,\\
\hphantom{$^{\dag^2}$}~F-69622-Villeurbanne, France}
\EmailDD{\href{mailto:gieres@ipnl.in2p3.fr}{gieres@ipnl.in2p3.fr}, \href{mailto:s.hohenegger@ipnl.in2p3.fr}{s.hohenegger@ipnl.in2p3.fr}}

\Address{$^{\dag^3}$~Deceased}

\Address{$^{\dag^4}$~Austro-Ukrainian Institute for Science and Technology c/o AUI, ITP,\\
\hphantom{$^{\dag^4}$}~Vienna University of Technology, Wiedner Hauptstra{\ss}e 8-10, A-1040 Vienna, Austria}
\EmailDD{\href{mailto:miw@hep.itp.tuwien.ac.at}{miw@hep.itp.tuwien.ac.at}}

\ArticleDates{Received September 19, 2018, in final form December 11, 2018; Published online December 23, 2018}

\Abstract{We discuss the formulation of classical field theoretical models on $n$-dimensional noncommutative space-time defined by a generic associative star product. A simple procedure for deriving conservation laws is presented and applied to field theories in noncommutative space-time to obtain local conservation laws (for the electric charge and for the energy-momentum tensor of free fields) and more generally an energy-momentum balance equation for interacting fields. For free field models an analogy with the damped harmonic oscillator in classical mechanics is pointed out, which allows us to get a physical understanding for the obtained conservation laws. To conclude, the formulation of field theories on curved noncommutative space is addressed.}


\Classification{81T75; 70H33; 81T13; 51P05; 81T75}

\begin{flushright}\begin{minipage}{140mm}
\it Manfred Schweda passed away in 2017 before the present work which he initiated was finished. His coauthors dedicate this paper to his memory with deep gratitude for his enthusiastic collaboration and friendship as well as for the example he has given to all of us over many years.
\end{minipage}
\end{flushright}

\section{Introduction}

Over the last twenty years a great amount of work has been devoted to the study of structural aspects and phenomenological applications of field theories on the simplest quantized space, namely the Groenewold--Moyal (or $\th$-deformed) space~\cite{Groenewold:1946,Moyal:1949}, e.g., see~\cite{Aschieri:2009zz,Banerjee:2009gr,Blaschke:2010kw, Grosse:2007jr,Rivasseau:2007a,Szabo:2001, Szabo:2006wx, Wallet:2007em,Wohlgenannt:2006dx, Wulkenhaar:2006si} and references therein for a review. On this space the theories are formulated in terms of ordinary functions by means of a deformed associative product, the so-called Groenewold--Moyal star product
\begin{gather}\label{eq:GMproduct}
(f\star g) (x) \equiv \re^{\frac{\ri}2 \th^{\mu\nu} \partial^x_\mu \partial^y_\nu } f(x) g(y) \big|_{y= x}, \qquad \mbox{i.e.,} \qquad
f\star g = fg + \tfrac {\ri}{2} \theta^{\mu\nu} \partial_\mu f \partial_\nu g + \mathcal O\big(\theta^2\big) .
\end{gather}
Here, the noncommutativity parameters $\theta^{\mu\nu} = - \theta^{\nu\mu} $ are real constants and expression~\eqref{eq:GMproduct} implies that the space-time coordinates $x^\m$ fulfill a Heisenberg-type algebra,
\begin{gather}\label{eq:CanDeform}
\starco{x^\mu}{x^\nu} \equiv x^{\mu} \star x^{\nu} - x^{\nu} \star x^{\mu} = \ri\th^{\mu\nu} .
\end{gather}
One refers to this case as the \emph{canonical} deformation. Quite generally, the interest in this and more general deformed spaces was triggered by their link with quantum gravity, quantum geo\-metry, string theory and
 $D$-branes, matrix models, the quantum Hall effect as well as other physical systems (see, e.g., \cite{Aschieri:2009zz,Banerjee:2009gr,Blaschke:2010kw, Deriglazov:2017jub, Grosse:2007jr,Horvathy:2010wv, Rivasseau:2007a,Szabo:2001, Szabo:2006wx, Wallet:2007em,Wohlgenannt:2006dx, Wulkenhaar:2006si} and references therein). Moreover, it was realized that quantum field theory is in some ways better behaved on noncommutative space-time than in ordinary space-time, e.g., see~\cite{Rivasseau:2007a} for an assessment.

In the present work, we are interested in classical field theories defined on a space-time for which the noncommutativity parameters $\theta^{\mu\nu}$ appearing in the algebra~\eqref{eq:CanDeform} are space-time dependent. These models, which appear to be more natural from the point of view of gravity and which may have some interesting physical applications, have been much less investigated in the literature\footnote{For different approaches to field theories on spaces with coordinate dependent noncommutativity in various dimensions, we refer to the works~\cite{Behr:2003qc, Behr:2003hg,Calmet:2003jv, Das:2003kw, Fosco:2004jb, Gayral:2005ih,Madore:2000en, Steinacker:2010rh}. For instance, the authors of~\cite{Gayral:2005ih} applied Rieffel's deformation quantization theory to Euclidean $\br ^4$ with a topology $\br^2 \times \br^2$:
This approach has the advantage of being non-perturbative in $(\theta^{\m \n})$, but it only works for a matrix $(\theta^{\m \n}(x))$ whose canonical form solely involves two non-vanishing entries $\theta^{34} = - \theta^{43} = \vartheta \big(x^1, x^2\big)$ where $\vartheta$ denotes an arbitrary smooth, positive, bounded function.}. Our work is based to a large extent on the mathematical results~\cite{Kupriyanov:2012rf, Kupriyanov:2008dn} obtained by V.G.~Kupriyanov (and his collaborators) and applied mostly in the context of quantum mechanics on noncommutative space~\cite{Gomes:2009tk,Gomes:2009rz,Kupriyanov:2012nb,Kupriyanov:2013jka,Kupriyanov:2015uxa}.
We should also mention the special instances where the noncommutativity parameters are linear in the coordinates (the so-called \emph{linear} or \emph{Lie algebra} case which is related to fuzzy spaces and $\kappa$-deformations) and the case where they are quadratic in the coordinates which is related to quantum groups, both cases having been the subject of various studies in the literature~-- see the reviews mentioned above
as well as~\cite{Kupriyanov:2015uxa}.

A basic issue in the construction of Lagrangian models in classical field theory consists in the study
of the underlying symmetries and of the conservation laws in differential or integral form.
These questions which are related to Noether's first theorem have been investigated in the case of constant noncommutativity parameters
by numerous authors, e.g., see~\cite{Balasin:2015hna, Geloun:2007zza, Schweda:2000,Micu:2000,Pengpan:2000kd}
and references therein. Already for this simplest instance of a
noncommutative space, technical complications arise, in particular
for the energy-momentum tensor in gauge field theories~-- see~\cite{Balasin:2015hna}.
The main concern of the present work is to address these questions in the simplest possible manner for a space-time with $x$-dependent
noncommutativity, both for free and interacting classical field theories.
Thus, we formulate a simple and general procedure for deriving local conservation laws (or balance equations)
which is based on the equations of motion and which follows the familiar line of reasoning in classical or quantum mechanics.
For bosonic and fermionic matter fields, we then derive the local form of charge conservation laws and of energy-momentum conservation laws
for free fields as well the energy-momentum balance equation for interacting fields. For those models which have already been discussed
in the context of quantum mechanics~\cite{Kupriyanov:2013jka} or of field theory on a space with a noncommutativity of Lie algebra-type~\cite{Kupriyanov:2015uxa},
we recover the same results.

The text is organized as follows.
In Section~\ref{sec:ncspace}, we present the set-up of noncommutative space-time defined by a generic
associative star product
as well as the characteristics of the latter.
In Section~\ref{sec:FTNC}, we discuss the formulation of classical field theory on such a space and we point out an analogy
of the free field models with the damped harmonic oscillator in classical mechanics.
 In Section~\ref{sec:conslaws}, we consider the derivation of conservation laws
in non-relativistic classical and quantum mechanics as a motivation for a quite
simple derivation of local conservation laws
in relativistic field theory. This procedure (relying on the equations of motion) as well as the properties of the star product
then allow for a straightforward derivation of local conservation laws in noncommutative space-time (along the lines followed
by V.G.~Kupriyanov for the Dirac equation
in noncommutative quantum mechanics).
In Section~\ref{sec:FTcurvedSpace}, we address the formulation of classical field theories on curved noncommutative space-time
by generalizing the description considered in flat space. More precisely, we follow the star product approach to gravity which
was recently put forward by M.~Dobrski~\cite{Dobrski:2015emm} and which appears to fit nicely into the framework discussed in flat space-time.
To conclude, we comment on the quantum theory in Section~\ref{sec:Conclusion}.
The aim of the appendices is to complement different parts of the text. In Appendix~\ref{sec:ConsStarProd},
we outline the passage from the operatorial approach
to the star product formulation as well as the construction of a closed star product.
We have deferred to Appendix~\ref{sec:MathComment} several mathematical comments on star products and on the deformed Leibniz rule,
as well as on two-dimensional space-time.
Appendix~\ref{sec:dampedosci} summarizes the Lagrangian and Hamiltonian formulations of the
damped harmonic oscillator in view of its similarities with free field theoretical models on noncommutative space-time.

\section{Noncommutative space}\label{sec:ncspace}

\paragraph{Operatorial approach:} We consider a deformation of $n$-dimensional Minkowski space (with metric signature $(+, - , \dots, -)$). More precisely, the space-time coordinates represent non\-com\-muting variables $\hat X ^\m$ (operators on some Hilbert space) whose noncommutativity is described by the relation
\begin{gather}\label{eq:one_operators}
\big[\hat X ^\mu , \hat{X}^\nu\big] = \ri \hat{\theta} ^{\mu\nu} \big(\hat X \big) ,
\end{gather}
where $\hat{\theta} ^{\mu\nu}$ is antisymmetric in its indices. For a general operator function $\hat{\theta} ^{\mu\nu}$ which is given as a power series in the variables $\hat X ^\rho$, an ordering prescription of operators must be defined, a natural choice being Weyl's symmetric ordering. The operator $ \hat{\theta} ^{\mu\nu} \big(\hat X\big)$ has a Weyl symbol~\cite{Berezin, Grensing:2013} whose leading order~\cite{Kupriyanov:2008dn}
will be denoted by $ \theta^{\mu\nu} (x)$. From the Jacobi identity for the algebra~\eqref{eq:one_operators} it follows~\cite{Kupriyanov:2012rf, Kupriyanov:2008dn} that the antisymmetric field $(\theta^{\mu\nu})$
has to satisfy the Jacobi identity for the Poisson bracket $\{ f, g \} \equiv {\theta} ^{\mu\nu} \pa_\mu f \pa_\n g$, i.e., the partial differential equations $0= \theta^{\mu\nu} \pa_\nu \theta^{\rho \sigma} +$
cyclic permutations of $\mu$, $\rho$, $\sigma$ (which we will refer to as the Poisson--Jacobi identity). Here and in the following, the derivatives are supposed to act only on the first factor which follows, i.e., $\partial_\mu f \partial_\nu g \equiv (\partial_\mu f )( \partial_\nu g)$.

In the present paper we do not follow the operatorial approach based on the commutation relations~\eqref{eq:one_operators}, but rather the equivalent star product approach which we will outline in the sequel (see~\cite{Bayen:1977ha, Dito:2002dr, Fedosov:1996, Kontsevich:1997vb} for the pioneering work, \cite{Esposito:2015, Waldmann:2007} for an introduction to the general theory, and~\cite{Behr:2003qc,Behr:2003hg,Calmet:2003jv,Kupriyanov:2012rf,Kupriyanov:2008dn} for an explicit and constructive approach). However, the operatorial approach is quite useful for describing some basic aspects and for a simple construction of the star product as we outline in Appendix~\ref{sec:ConsStarProd}.

\paragraph{Star product approach:} An equivalent description of the deformed space above can be given by passing (by virtue of an isomorphic map) from the noncommutative algebra generated by the operators $\hat X ^\m$ to the
commutative, associative algebra $C^{\infty} (\br ^n)$ of smooth functions (depending on real variables~$x^\m$) equipped with an additional so-called star product~$\star$ which is associative, but noncommutative:
The \emph{star-commutator} is then given by
\begin{gather}\label{eq:one}
\big[x^\mu\stackrel{\star}, x^\nu\big] = \ri \theta^{\mu\nu} (x) ,
\end{gather}
where the real-valued function $\theta^{\mu\nu}$ with $\theta^{\mu\nu} = - \theta^{\nu\mu} $ is the afore-mentioned symbol of the ope\-rator $\hat{\theta} ^{\mu\nu} (\hat{X} )$ which satisfies the Poisson--Jacobi identity. Relation~\eqref{eq:one} for the variables $x^\m$ is satisfied if the ordinary product $fg$ (of any two smooth functions~$f$, $g$ of the variables $x^\m$) is deformed according to
\begin{gather}
f\star g = fg + \tfrac {\ri}{2} \theta^{\mu\nu} \partial_\mu f \partial_\nu g + \mathcal O\big(\theta^2\big) , \qquad \mbox{hence} \qquad
[f \stackrel{\star}{,} g] = \ri \theta^{\mu\nu} \partial_\mu f \partial_\nu g + \mathcal O\big(\theta^2\big) .\label{eq:3.2}
\end{gather}
As is customary in the theory of formal star products or, more generally, in Gerstenhaber's theory of algebraic deformations~\cite{Waldmann:2007}, a~constant real formal deformation parameter $h$ can be factored out of $\theta^{\m \n}$ by writing $\theta^{\m \n} = h \Theta^{\m \n}$ so that expression~\eqref{eq:3.2} may be viewed as a formal power series in~$h$. By a slight abuse of terminology, we refer to the order of $h$ in the series expansion as the \emph{order of $\theta$}.

The correspondence between the operator and star product formulations is such that we have (in terms of the notation $\hat W [f] \equiv \hat f \big(\hat X\big)$ for the operators)
\begin{gather}\label{eq:Weylcorres}
\hat W [f \star g ] = \hat W [f] \circ \hat W [g] ,
\end{gather}
where $\circ$ denotes the product of operators. The operator $\hat W [f] \equiv \hat f \big(\hat X\big)$ acts on the unit function~$1$ according to $ \hat f \big(\hat X\big) 1 = f$, hence relation~\eqref{eq:Weylcorres}
implies that its action on a smooth function~$g$ is given by $\hat f \big(\hat X\big) g = f \star g$. According to this equation, an explicit expression for $f \star g$ can be determined perturbatively~\cite{Kupriyanov:2008dn} by looking for an expansion of $\hat X ^\m$ (and thereby of the Weyl ordered function $\hat f \big(\hat X\big)$) as a differential polynomial in the standard, commuting position ope\-ra\-tors~$X^\m$ and in~$\pa_\n$ (see Appendix~\ref{sec:ConsStarProd}): more precisely, one may easily check that the expansion $\hat X ^\m = X^\m + \frac{\ri}{2} \theta^{\m \n} \pa_\n + {\cal O}\big(\theta^2\big)$ (which is familiar from the definition of the Groenewold--Moyal product) satisfies the commutation relation~\eqref{eq:one_operators} and that equation $\hat f \big(\hat X\big) g = f \star g$ yields the expansion~\eqref{eq:3.2} to first order in $\theta$. The expression that one finds~\cite{Kupriyanov:2008dn} for the star product to second order in $\theta$ is given by
\begin{gather}
f\star g = fg + \tfrac {\ri}{2} \theta^{\mu\nu} \partial_\mu f \partial_\nu g - \tfrac{1}{8} \theta^{\rho \s} \theta^{\m \n} \pa_\rho \pa_\m f \pa_\s \pa_\n g \nonumber\\
\hphantom{f\star g =}{} - \tfrac{1}{12} \theta^{\rho \s} \pa_\s \theta^{\m \n} ( \pa_\rho \pa_\m f \pa_\n g - \pa_\m f \pa_\rho \pa_\n g) + \mathcal O\big(\theta^3\big) .\label{eq:KontStarProduct}
\end{gather}
This result (which is referred to as the \emph{Weyl star product}~\cite{Kupriyanov:2015uxa}) coincides with the explicit expression which has been given for $\br ^n$ by Kontsevich up to second order at the beginning of his seminal article~\cite{Kontsevich:1997vb} (e.g., see~\cite{Esposito:2015} for a nice introductory review of Kontsevich's work). For constant noncommutativity parameters, the star product~\eqref{eq:KontStarProduct} reduces to the Groenewold--Moyal star product at the given order. The first three terms may be referred to as the ``Groenewold--Moyal-like part''~\cite{Herbst:2001ai} since they have the same form as the Groenewold--Moyal
 star product up to this order. Quite generally the terms of even order in $\theta$ are symmetric with respect to the exchange of the functions~$f$ and~$g$ while the odd order terms are antisymmetric~\cite{Kupriyanov:2008dn, Kupriyanov:2015uxa}. The series~\eqref{eq:KontStarProduct} can be interpreted physically as the perturbative expansion of the path integral for a non-linear sigma model for the world-sheet description of bosonic strings~\cite{Cattaneo:1999fm, Cornalba:2001sm, Herbst:2001ai, Szabo:2006wx}.

\paragraph{Closed star product:} For the formulation of quantum mechanics or field theory we are interested in so-called \emph{closed star products}~\cite{ConnesFlatoSternheimer}, i.e., star products for which the integral of a~function $f$ over $\br ^n$ is the trace of the operator $\hat f \big(\hat X\big)$:
\begin{gather*}
\Tr \hat f \big(\hat X\big) = \int_{\br ^n} {\rm d}^nx w(x) f(x) .
\end{gather*}
Here, the standard integration measure ${\rm d}^nx$ has been modified\footnote{This modification and the expression~\eqref{eq:w} below for~$w$ are natural for symplectic manifolds and have been noted in the case of $\br^n$ by different authors~\cite{Calmet:2003jv,Felder:2000nc, Fosco:2004jb, Steinacker:2010rh}.} by a weight factor $w$ so as to ensure the cyclicity of the trace, i.e., the validity of the so-called \emph{closedness} or \emph{closure relation}
\begin{gather}
 \int_{\br ^n} {\rm d}^nx w (f \star g ) = \int_{\br ^n} {\rm d}^nx w f g.\label{eq:Closure}
\end{gather}
More precisely, substitution of the expansion~\eqref{eq:KontStarProduct} into relation~\eqref{eq:Closure} and the assumption that~$f$ and~$g$ are smooth functions of fast decay at infinity with respect to the considered integration measure (so that all boundary terms in the integral vanish) yield
\begin{gather}\label{eq:ClosRelOrder1}
0= \int_{\br ^n} {\rm d}^nx w ( f \star g - fg ) = - \tfrac{\ri}{2} \int_{\br ^n} {\rm d}^nx \pa_\m ( w \theta^{\m \n} ) f \pa_\n g + {\cal O} \big(\theta^2\big) .
 \end{gather}
Since this condition has to hold for arbitrary functions $f$, $g$, we obtain the condition
\begin{gather}
\pa_\m (w \theta^{\m \n} ) =0 , \qquad \mbox{i.e.,} \qquad w \partial_\mu \theta^{\mu\nu} = - \partial_\mu w \theta^{\mu\nu} .\label{eq:PDEw}
\end{gather}
This relation means that \emph{the tensor $(\theta ^{\m \n} )$ is divergenceless} with respect to the integration measure $w {\rm d}^nx$. (We note that for a curved $n$-brane in a flat background space-time, the divergence condition~\eqref{eq:PDEw} admits the physical interpretation of a Born--Infeld equation of motion on the brane~\cite{Szabo:2006wx}.) Thus, for non-constant functions $\theta^{\m \n}$, the factor $w$ cannot be chosen to be a~constant. Given the noncommutativity parameters $\theta^{\m \n} (x) $, this partial differential equation determines the function~$w$ in terms of $\theta^{\m \n}$. By using the usual formula for the infinitesimal
variation of the determinant of a matrix as well as the Poisson--Jacobi identity, one can readily check that a~solution~\cite{Behr:2003hg, Kupriyanov:2012rf, Kupriyanov:2013jka} of equation~\eqref{eq:PDEw} is given as follows if the matrix $(\theta ^{\m \n} (x))$ is \emph{invertible} for all $x \in \br ^n$ (which requires $n$ to be even):
\begin{gather}\label{eq:w}
w = \big( \det (\theta^{\m \n} ) \big)^{-1/2} .
\end{gather}
In this respect, we recall that the determinant of an invertible, real, antisymmetric matrix $\Theta \equiv (\theta ^{\m \n} )$ of even order has a strictly positive determinant; in fact~\cite{Grensing:2013}, this determinant is the square of the so-called \emph{Pfaffian} of the matrix, $\det \Theta = ( \operatorname{Pf} \Theta )^2$.

If the rank of the matrix $(\theta ^{\m \n} )$ is not even and maximal for all $x\in\mathbb{R}^n$, then one has to look for the corresponding solutions of equation~\eqref{eq:PDEw} (e.g., see~\cite{Fosco:2004jb} for the two-dimensional case). In the following we will assume that the matrix $(\theta ^{\m \n} )$ is invertible whenever needed, the weight function $w$ being then given by expression~\eqref{eq:w}. (The latter assumption is mainly
made for convenience: apart from Section~\ref{sec:FTcurvedSpace} below, all we really need is the existence of a non-vanishing weight factor~$w$ satisfying equation~\eqref{eq:PDEw}.) The quadratic terms in the closure relation~\eqref{eq:ClosRelOrder1} are discussed in Appendix~\ref{sec:ConsStarProd} and we will address them in equations~\eqref{eq:gaugeTransfStar}, \eqref{eq:StarProduct} below.

\paragraph{Some mathematical remarks:} The antisymmetric tensor $\theta^{\m \n}$ satisfying the Poisson--Jacobi identity represents a \emph{Poisson tensor} (also referred to as \emph{Poisson bivector field}~\cite{Waldmann:2007}) and relation~\eqref{eq:3.2} identifies the star-commutator as a deformation of the Poisson bracket. If we assume that the matrix $(\theta^{\m \n})$ is invertible at all points, then it admits an inverse matrix $(\omega_{\m \n}) \equiv (\theta^{\m \n}) ^{-1}$. The latter matrix is non-degenerate, antisymmetric and, by virtue of the Poisson--Jacobi identity for $\theta^{\m \n}$, its components satisfy the relation
$0= \pa_\rho \omega_{\m \n} +$ cyclic permutations of the indices. Thus $ \underline{\omega} \equiv \frac{1}{2} \omega _{\m \n} {\rm d}x^\m \wedge {\rm d}x^\n$ is a \emph{symplectic two-form} and we have $w = \sqrt{\det (\omega_{\m \n} )} $. We note that the latter factor is reminiscent of the density $\sqrt{g} \equiv \sqrt{ \det (g_{\m \n}) }$ which appears for the integration on a~Riemannian manifold with metric tensor $(g_{\m \n} )$. Indeed, the \emph{canonical volume form} (or so-called \emph{Liouville measure}) on a~symplectic manifold of dimension $n=2m$ reads~\cite{Bayen:1977ha, Fedosov:1996, Waldmann:2007}
\begin{gather}\label{eq:VolElem}
{\rm d}V \equiv \frac{1}{m!} \underline{\omega}^{m} = \operatorname{Pf} (\omega_{\m \n} ) {\rm d}^n x = \sqrt{ \det (\omega_{\m \n} ) } {\rm d}^n x = w {\rm d}^nx ,
\end{gather}
where we did not spell out the exterior product symbols. For a symplectic manifold $M$, the integral $\int_{M} {\rm d}^n x w f$ of a function $f\colon M\to \mathbb{C}$ with respect to the Liouville measure is also qualified as a \emph{Poisson trace}~\cite{Waldmann:2007}.

Concerning our considerations on star products, we emphasize that Minkowski space $\br ^n$ is considered (for $n$ even) as a~\emph{symplectic manifold} or as a \emph{Poisson manifold} rather than a~symplectic vector space~\cite{Marsden:1999} or a Poisson algebra~\cite{PoissonBible} since the tensor $(\theta^{\m \n })$ is not constant. Thus, in general $(\omega_{\m \n} ) $ and $(\theta^{\m \n })$ cannot be cast into canonical form (Darboux or canonical coordinates) by a \emph{linear} change of coordinates: This contrasts the case of constant deformation parameters (Groenewold--Moyal star product). As for the star products, they are to be viewed as non-trivial associative deformations of the associative algebra $C^{\infty}(\br ^n )$~\cite{Waldmann:2007}. For the case of a~general Poisson manifold $M$ (generalizing $M=\br ^n$), relation~\eqref{eq:Closure} is known as the \emph{generalized Connes--Flato--Sternheimer conjecture}. In this respect we note that the authors of~\cite{Felder:2000nc} have shown that for any Poisson tensor which is divergenceless with respect to some volume form,
there exists a star product which satisfies \eqref{eq:Closure}.

\paragraph{Explicit expression of closed star product:} Due to the non-trivial weight factor $w$ which has to be present in equation~\eqref{eq:Closure}, the terms of order $\theta^2$ in the Weyl star product~\eqref{eq:KontStarProduct} do not satisfy the closure relation. However, as pointed out in~\cite{Kupriyanov:2012rf} and as outlined in Appendix~\ref{sec:ConsStarProd}, one can pass over to a \emph{gauge
equivalent}~\cite{Kontsevich:1997vb} \emph{star product} $\star '$ which satisfies this relation to order~$\theta^2$. The corresponding gauge transformation of the Weyl star product is given by
\begin{gather}\label{eq:gaugeTransfStar}
f \star g \longmapsto f \star ' g \equiv D^{-1} ( Df \star Dg ) , \qquad \mbox{with} \qquad D = \id + \tfrac{1}{48 w} \pa_\m ( w \theta^{\rho \s} \pa_\s \theta^{\m \n} ) \pa_\rho \pa_\n ,
\end{gather}
and it readily leads to the following expression to order $\theta^2$ of a closed star product~\cite{Kupriyanov:2012rf} (in which we dropped the prime on $\star '$):
\begin{gather} 
f\star g = fg + \tfrac {\ri}{2} \theta^{\mu\nu} \partial_\mu f \partial_\nu g - \tfrac{1}{8} \theta^{\rho \s} \theta^{\m \n} \pa_\rho \pa_\m f \pa_\s \pa_\n g
 - \tfrac{1}{12} \theta^{\rho \s} \pa_\s \theta^{\m \n} ( \pa_\rho \pa_\m f \pa_\n g - \pa_\m f \pa_\rho \pa_\n g)\nonumber\\
\hphantom{f\star g =}{} - \tfrac{1}{24 w} \pa_\m ( w \theta^{\rho \s} \pa_\s \theta^{\m \n} ) \pa_\rho f \pa_\n g + \mathcal O\big(\theta^3\big) .\label{eq:StarProduct}
\end{gather}
In summary, for a given antisymmetric tensor field $(\theta ^{\m \n})$ satisfying the Poisson--Jacobi identity, the star product~\eqref{eq:StarProduct} satisfies the closure relation~\eqref{eq:Closure} to order~$\theta^2$ with an integration measure $w {\rm d}^nx$ with respect to which the Poisson tensor $(\theta ^{\m \n})$ is divergenceless.

By construction, the expressions $w f\star g$ and $w fg $ only differ by a total derivative so that equality~\eqref{eq:Closure} holds. For later reference, we spell out the explicit expression of this derivative~\cite{Kupriyanov:2013jka} (for which the Jacobi identity has again been used):
\begin{gather}\label{eq:localClosure}
w (f\star g) = w fg + \pa_\rho a^{\rho} (f,g) ,
\end{gather}
with
\begin{gather*}
 a^{\rho} (f,g) = w \big[ \tfrac{\ri}{4} \theta^{\rho \s} ( f \pa_\s g - \pa_\s f g)
 + \tfrac{1}{16} \theta^{\rho \s} \theta^{\m \n} ( \pa_\s \pa_\m f \pa_\n g - \pa_\m f \pa_\s \pa_\n g)\nonumber \\
\hphantom{a^{\rho} (f,g) =}{} + \tfrac{1}{48} ( \theta^{\n \s} \pa_\s \theta^{\m \rho} - \theta^{\m \s} \pa_\s \theta^{\rho \n} ) \pa_\m f \pa_\n g \big] + \mathcal O\big(\theta^3\big) .
\end{gather*}
In the case of constant noncommutativity parameters $\theta^{\m \n}$, the star operation is not $x$-dependent in the sense that the Leibniz rule $\pa_\m (f \star g ) = \pa_\m f \star g + f \star \pa_\m g$ holds. This is no longer true for non-constant functions $\theta^{\m \n}$ where one has a modified rule for the differentiation, see equations~\eqref{eq:DerStarProd}--\eqref{eq:IdSP} below.

It is worthwhile to note the behavior of the star product of complex-valued functions under the operation of complex conjugation~\cite{Kupriyanov:2013jka}:
\begin{gather*}
(f \star g )^* = g^* \star f^* .
\end{gather*}
This relation can be explicitly checked for the expansion~\eqref{eq:StarProduct}, the verification for the last term making use of the Jacobi identity and of relation~\eqref{eq:PDEw} which is satisfied by the function $w$.

We refer to~Appendix~\ref{sec:MathComment} for some mathematical comments on the potential relationship between the star products considered here and the approach of A.~Connes to noncommutative geometry.

\paragraph{Some particular cases:} As pointed out in~\cite{Kupriyanov:2008dn}, some simplifications in the perturbative expansions underlying the star products occur for the special case of \emph{linear} Poisson structures,
i.e., for a commutator algebra of Lie algebra type, $\big[\hat X^\m , \hat X^\n \big] = \ri f^{\m \n}_{\rho} \hat X^\rho$. The latter instance (and in particular the case of the Lie algebra ${\bf su} (2)$ where $\big[\hat X^i , \hat X^j \big] = \ri \varepsilon^{ijk} \hat X^k$ and where the weight function~$w$ can be chosen to be constant, $w\equiv 1$, since $\pa_i \theta^{ij} = \pa_i \big(\varepsilon^{ijk} x^k\big) =0$) has been studied in detail
in~\cite{Kupriyanov:2015uxa} (see also~\cite{Juric:2016cfp} for some elaborations and applications). Remarkably, some closed expressions can then be determined in the case of ${\bf su} (2)$ (and even for more general Lie algebras) for the operators $\hat X ^i$, for the star product as well as for the gauge transformation relating the Weyl star product and the closed star product. Moreover, the latter product can be identified~\cite{Dito:2002dr, Kupriyanov:2015uxa} with the one following from the so-called Duflo quantization map for ${\bf su} (2)$ which appears to be~\cite{Kupriyanov:2015uxa} the mathematically preferred quantization in this context (see also~\cite{Rosa:2012pr}).

We will briefly come back to the ${\bf su} (2)$ case in the concluding Section~\ref{sec:Conclusion} with some comments on the properties of quantum field theories on such spaces. The case of two space-time dimensions, which also represents an instance of particular interest, is commented upon in~Appendix~\ref{sec:MathComment}.

\section{Field theory on noncommutative space}\label{sec:FTNC}

\paragraph{Generalities:} In field theory the action functionals are generally expressed in terms of an $L^2$-type scalar product $\langle \cdot, \cdot \rangle$ of complex-valued fields~$f$,~$g$. In the present context, this scalar product involves the weight factor $w = 1/ \sqrt{\det (\theta^{\m \n} )} $ (the latter being strictly positive with the assumption that the matrix $(\theta^{\m \n}(x))$ is non-singular for all $x\in \br ^n$):
\begin{gather}\label{eq:defSP}
\langle f, g \rangle \equiv \int_{\br ^n} {\rm d}^n x w f^* g .
\end{gather}
Any derivatives appearing in the Lagrangian should represent anti-Hermitian operators with respect to this scalar product, very much as the momentum operator in relativistic quantum mechanics should be Hermitian with respect to the scalar product of wave functions~$f$,~$g$. As remarked by Kupriyanov~\cite{Kupriyanov:2013jka} following the works~\cite{Bagchi:2009wb, Fring:2010pw, Gomes:2009tk}, this implies that the momentum operator in relativistic quantum mechanics is no longer given by $p_\m \equiv -\ri \pa_\m$, but presently involves an additional term, $\hat p_\m \equiv - \ri \pa_\m -\ri \pa_\m \big(\ln w^{1/2} \big)$, so as to have
 $\langle f, \hat p_\m g \rangle = \langle \hat p_\m f,g \rangle$ for all wave functions $f$,~$g$ belonging to the domain of definition of the operator\footnote{This is reminiscent of the introduction and form of the Newton--Wigner position operator for the Klein--Gordon wave equation in relativistic quantum mechanics~\cite{Schweber}. Indeed, for the positive energy solutions of the Klein--Gordon equation in~$\br ^4$, the scalar product of momentum space wave functions $\varphi$, $\psi$ reads $\langle \varphi , \psi \rangle = \int_{\br ^3} \frac{{\rm d}^3p}{\omega (\vec p ) } \overline{{\varphi } (\vec p )} \psi (\vec p )$ where ${{\rm d}^3p}/\omega (\vec p )$ with $\omega (\vec p ) \equiv \sqrt{{\vec p}\,{}^{ 2} +m^2}$ represents the Lorentz invariant integration measure over the mass hyperboloid $p^2 = m^2$. The usual expression for the position operator in momentum space, i.e., $\vec X \equiv \ri \vec{\pa}_{\vec p} \equiv \ri \vec{\pa} / \pa \vec p$, is not Hermitian with respect to the given scalar product, but the Newton--Wigner position operator is and reads $\hat{\vec{X}} \equiv \ri \big[ \vec{\pa}_{\vec p} + \vec{\pa}_{\vec p} \big(\ln \omega^{-1/2} \big) \big]= \ri \big[ \vec{\pa}_{\vec p} -\frac{1}{2} \frac{\vec p }{{\vec p}^{ 2} +m^2} \big]$. } $\hat p_\m $. We note that
 \begin{gather} \label{eq:defCD}
 \hat p_\m f = - \ri \cd_\m f , \qquad \mbox{with} \qquad
 \cd_\m f \equiv \big[ \pa_\m + \pa_\m \big(\ln w^{1/2}\big) \big] f = w^{-1/2} \pa_\m \big( w^{1/2} f\big) ,
 \end{gather}
 and $\langle f, \cd_\m g \rangle = - \langle \cd_\m f,g \rangle$ for smooth functions $f$, $g$. Here, the derivative $\cd_\m$ has the form of a~covariant derivative in Abelian gauge theory with the particular gauge potential
 \begin{gather} \label{eq:PotA}
A_\m \equiv \pa_\m \big(\ln w^{1/2}\big) .
\end{gather}
Since the latter is pure gauge, the covariant derivatives commute with each other (i.e., $[ \cd_\m , \cd_\n ] =0$) and they are related by a gauge transformation to the ordinary derivatives (namely $\cd_\m = w^{-1/2} \circ \pa_\m \circ w^{1/2}$). Actually, the fact that the coefficient $w = 1/\sqrt{ \det (\theta^{\m \n}) } = \sqrt{ \det (\omega_{\m \n})} $ resembles the metric coefficient $\sqrt{|g|} \equiv | \det (g_{\m \n} )|^{1/2}$ in general relativity, indicates that the gauge potential~\eqref{eq:PotA} may be viewed as the analogue of the contraction $\Gamma^{\n}_{\m \n} = \pa_\m \big(\ln \sqrt{|g|} \big)$ of the Christoffel symbols $\Gamma^{\lambda}_{\m \n}$
which appear in the covariant derivatives on a pseudo-Riemannian manifold (equipped with the Levi-Civita connection).

Before considering some field theoretical models, we mention a general identity~\cite{Kupriyanov:2013jka} for star products involving derivatives $\cd_\m $ which is useful for the derivation of conservation laws in field theory. By differentiating equation~\eqref{eq:localClosure} with respect to $x^\m$ we obtain
\begin{gather}\label{eq:DerStarProd}
\pa_\m [ w (f\star g) ] = w [ \cd_\m f g + f \cd_\m g ] + \pa_\rho \pa_\m a^{\rho} (f,g) .
\end{gather}
If we now apply equation~\eqref{eq:localClosure} to the first contribution on the right hand side, we conclude that
\begin{gather}\label{eq:IdSP}
w [ \cd _\m f \star g + f \star \cd _\m g ] = \pa_\m [ w (f\star g) ] + \pa_\rho b^{\rho}_\m (f,g) ,
\end{gather}
where the last term is given by
\begin{gather*}
 b^{\rho}_\m (f,g) \equiv a^{\rho} (\cd _\m f,g)+ a^{\rho} (f, \cd _\m g) - \pa_\m a^{\rho} ( f,g)
 = \tfrac{\ri}{4} w \pa_\m \theta^{\rho \s} ( \pa_\s f g - f \pa_\s g) + \mathcal O\big(\theta^2\big) .
\end{gather*}
Here, the Poisson--Jacobi identity was again taken into account for deriving the last expression.

The derivative~\eqref{eq:DerStarProd} can also be expanded by using the Leibniz rule for partial derivatives:
\begin{gather}\label{ex:ExpDerivStar}
\pa_\m [ w (f\star g) ] = w [ \pa_\m + \pa_\m (\ln w)] (f\star g) .
\end{gather}
Comparison of this result with expression~\eqref{eq:defCD}, i.e., $\cd_\m f \equiv \big[ \pa_\m + \frac{1}{2} \pa_\m (\ln w) \big] f$, then suggests to attribute a \emph{weight} (or \emph{degree} or \emph{charge}) $1/2$ to a function~$f$ and a weight $1$ to the star product of two such functions, so that~\eqref{ex:ExpDerivStar} represents the covariant derivative of $f \star g$. With this notation, relation~\eqref{eq:IdSP} reads
\begin{gather}\label{eq:LeibnizRule}
\pa_\m [ w (f\star g) ] = w \cd_\m (f\star g) = w [ \cd _\m f \star g + f \star \cd _\m g ] - \pa_\rho b^{\rho}_\m (f,g) ,
\end{gather}
i.e., the \emph{Leibniz rule} for covariant derivatives with a correction term $\pa_\rho b^{\rho}_\m (f,g)$. Some mathematical considerations concerning the deformed Leibniz rule~\eqref{eq:LeibnizRule} are presented in~Appendix~\ref{sec:MathComment}. Here, we only note that some alternative approaches are based on the introduction of derivatives~$D_\m$ which satisfy the Leibniz rule, but which do not commute with each other and which generally represent infinite power series in the elements of the matrix $(\theta^{\m \n})$ and of its inverse, e.g., see~\cite{Fosco:2004jb}. The Leibniz rule can also be imposed within a more abstract approach based on Hopf algebras and Drinfeld twists or by following an approach based on $L_{\infty}$-algebras, i.e., generalized DGLA's (differential graded Lie algebras) with a `mild' violation of associativity, see~\cite{Blumenhagen:2018kwq}.

\paragraph{Lagrangian models:} With the ingredients introduced above, field theoretical Lagrangian models on noncommutative space with a given $\theta$-tensor can now be defined by starting with models on ordinary space,
replacing the integration measure $\int_{\br ^n} {\rm d}^n x $ by $\int_{\br ^n} {\rm d}^n x w $, ordinary derivatives $\pa_\m$ by the covariant derivative $\cd_\m$ and ordinary products by star products.
For a~complex scalar field $\phi$ with a quadratic self-interaction we thus obtain the action functio\-nal~\cite{Kupriyanov:2012rf, Kupriyanov:2015uxa}
\begin{gather*}
S [\phi,\phi^*] = \langle \cd_\rho \phi^* , \cd^\rho \phi \rangle- m^2 \langle\phi^* , \phi \rangle- \frac{\cg}{2}\langle \phi^* \star \phi , \phi^* \star \phi \rangle ,
\end{gather*}
i.e.,\footnote{An alternative self-interaction~\cite{Arefeva:2000uu} given by the monomial $\phi^*\star\phi^*\star\phi\star\phi$ could be considered as well, but we will not discuss this other model here.}
\begin{gather}
S [\phi,\phi^*] =\int_{\br ^n} {\rm d}^n x w \left[\cd^\rho \phi^* \cd_\rho \phi - m^2 \phi^* \phi - \frac{\cg}{2} (\phi^* \star \phi)^2\right] , \label{eq:ActScalar}
\end{gather}
where we dropped one star in each term in accordance with the general property~\eqref{eq:Closure}. For the Dirac spinor $\psi$ coupled to an external $U (1)$-gauge field $(A^\m )$, we have the action~\cite{Kupriyanov:2013jka}
 \begin{gather}\label{eq:ActDirac}
S [\psi, \bar{\psi} ] = \int_{\br ^n} d^n x w \big[ \tfrac{\ri}{2} \bar{\psi} \gamma^\rho \cd _\rho \psi - \tfrac{\ri}{2} \cd _\rho \bar{\psi} \gamma^\rho \psi-m \bar{\psi} \psi+ e \bar{\psi} \gamma^\rho A_\rho \star \psi
\big] .
\end{gather}
For a $U(1)$-gauge field $(A^\m )$ in four dimensions, the action functional reads
\begin{gather}
 S[A]= -\tfrac{1}{4} \int_{\br ^4} {\rm d}^4 x w F_{\m\n}\star F^{\m\n} , \qquad
 \textrm{with} \qquad
 F_{\m\n} \equiv \cd_\m A_\n- \cd_\n A_\m-\ig \starco{A_\m}{A_\n} . \label{eq:gaugeaction}
\end{gather}

For an action functional $S$ depending on a bosonic field $\varphi $ (i.e., a scalar field $\phi$ or a gauge field $(A^\m )$), the functional derivative is defined by
\begin{gather*}
\delta S = \int_{\br ^n} {\rm d}^n x w \frac{\delta S}{\delta \varphi} \delta \varphi .
\end{gather*}
The components of a Dirac spinor $\psi$ are supposed to be anticommuting variables and the corresponding functional derivatives are defined by
\begin{gather*}
\delta S = \int_{\br ^n} {\rm d}^n x w \left[ \frac{\delta S}{\delta \psi} \delta \psi+\delta \bar{\psi} \frac{\delta S}{\delta \bar{\psi}}\right] .
\end{gather*}

The equations of motion associated to the action functionals~\eqref{eq:ActScalar}--\eqref{eq:gaugeaction} are respectively given by the vanishing of the following functional derivatives:
\begin{subequations}
\begin{gather}
-\frac{\delta S}{\delta \phi^*} = \big(\cd_\m \cd^\m + m^2\big) \phi + \cg (\phi \star \phi^*) \star \phi , \nonumber\\
-\frac{\delta S}{\delta \phi} = \big(\cd_\m \cd^\m + m^2\big) \phi^* + \cg (\phi^* \star \phi) \star \phi ^* ,\label{eq:FunctDerScal}\\
\label{eq:FunctDerDir}
\frac{\delta S }{\delta \bar{\psi}} = (\ri \gamma^\m \cd_\m -m ) \psi + e \gamma^\m A_\m \star \psi , \qquad
\frac{\delta S }{\delta {\psi}} = -\ri \cd _\m \bar{\psi} \gamma^\m -m \bar{\psi} + e \bar{\psi} \gamma^\m \star A_\m,\\
\label{eq:FunctDerGF}
\frac{\delta S}{\delta A_\n} = D_\m F^{\m \n} \equiv \cd_\m F^{\m \n} -\ri g\starco{A_\mu}{ F^{\m \n}} .
\end{gather}
\end{subequations}

Concerning the Dirac field action~\eqref{eq:ActDirac}, we recall that a classically equivalent (though non-real) expression for the kinetic term can be obtained by partial integration and presently reads $\int_{\br ^n} {\rm d}^n x w \big[ {\ri} \bar{\psi} \gamma^\rho \cd _\rho \psi \big]$.

Under an infinitesimal gauge transformation parametrized by a function $x \mapsto \alpha (x)$, the $U(1)$-gauge field $(A^\m )$ transforms with the gauge covariant derivative,
\begin{gather*}
\delta A_\m = D_\m \alpha = \cd_\m \alpha -\ri g \starco{A_\mu}{\alpha } .
\end{gather*}
By virtue of relation~\eqref{eq:LeibnizRule}, this induces the following transformation law of the field strength tensor $F_{\m \n}$:
\begin{gather*}
\delta F_{\m \n} = -\ri g\starco{F_{\m \n}}{\alpha } + \ri g w^{-1} \pa_\rho \big[b^{\rho} _\m (A_\n, \alpha ) - b^{\rho} _\m ( \alpha , A_\n ) - (\m \leftrightarrow \n ) \big] .
\end{gather*}
This variation leaves the gauge field action~\eqref{eq:gaugeaction} invariant. Concerning the field strength we remark that the modified Leibniz rule~\eqref{eq:LeibnizRule} for the derivatives $\cd_\mu$ implies a modified \emph{Bianchi identity:}
\begin{gather}\label{eq:BianchiId}
 D_\rho F_{\m \n} + \mbox{cyclic permut.} = \ri g w^{-1} \pa_\sigma \big[ b^\sigma_\rho (A_\m , A_\n ) - b^\sigma_\rho (A_\n , A_\m ) + \mbox{cyclic permut.} \big] .
\end{gather}
Since the Bianchi identity for the field strength generally reflects the Jacobi identity for the gauge covariant derivatives, the right hand side of equation~\eqref{eq:BianchiId} (which expresses the deviation from the Jacobi identity for star-commutators of gauge covariant derivatives) may be qualified as \emph{`gauge' Jacobiator}~\cite{Hohm:2017pnh, Waldmann:2007} for these commutators. The noncommutative $U(1)$-gauge theory as well as the Chern--Simons theory in three dimensions have been discussed in terms of $L_\infty$-algebras in~\cite{Blumenhagen:2018kwq}.

\paragraph{Limiting cases:} If the noncommutativity parameters $\theta^{\m \n}$ are constant, then the weight factor is constant too. Then, it can be factored out of the integrals~\eqref{eq:ActScalar}, \eqref{eq:ActDirac}, \eqref{eq:gaugeaction} and simply be dropped: thus we recover the canonical deformation case described by the Groenewold--Moyal star product. The commutative theory is obtained from the latter case (which does not involve~$w$ anymore) by letting the constant parameters $\theta^{\m \n}$ go to zero.

\paragraph{Free field models:} Let us have a closer look at the \emph{free} scalar and Dirac field actions. From the definition~\eqref{eq:defCD} of the covariant derivative, it follows that
\begin{gather}
S_{c=0} [\phi,\phi^*] = \int_{\br ^n} {\rm d}^n x \big[ \pa^\rho \Phi^* \pa_\rho \Phi - m^2 \Phi^* \Phi\big] , \qquad \mbox{with} \qquad \Phi \equiv \sqrt{w} \phi,\nonumber\\
S_{e=0} [\psi, \bar{\psi} ] = \int_{\br ^n}{\rm d}^n x \big[\tfrac{\ri}{2} \bar{\Psi} \gamma^\rho \pa _\rho \Psi - \tfrac{\ri}{2} \pa _\rho \bar{\Psi} \gamma^\rho \Psi -m \bar{\Psi} \Psi\big] , \qquad \mbox{with} \qquad
\Psi \equiv \sqrt{w} \psi .\label{eq:FreeModelActions}
\end{gather}
Thus, contrary to the case of constant noncommutativity parameters, the free field actions presently do not have the same form as in commutative space: they only do so after having expressed these actions in terms of the $\th$-dependent fields~$\Phi$,~$\Psi$. In terms of the original fields~$\phi$,~$\psi$, the free field equations have the form
\begin{gather}\label{eq:FreeFEphi}
0 = \frac{1}{\sqrt{w}} \pa_\m \pa^\m \big(\sqrt{w} \phi\big) + m^2 \phi , \qquad 0 = \frac{1}{\sqrt{w}} \ri \gamma^\m \pa_\m \big(\sqrt{w} \psi \big) -m \psi .
\end{gather}
These equations have a formal analogy with the equation of motion for a damped harmonic oscillator in non-relativistic mechanics (see Appendix~\ref{sec:dampedosci}): The latter equation reads
\begin{gather}
0 = \frac{1}{{\wt}} \pa_t (\wt \dot q ) + \omega^2 q , \qquad \mbox{with} \qquad \wt \equiv \re^{\gamma t}\nonumber\\
\hphantom{0}= \ddot q + \gamma \dot q + \omega^2 q ,\label{eq:EOMdampOsc}
\end{gather}
and can be rewritten in an undamped form, $\ddot Q + \Omega^2 Q =0$, by virtue of the redefinition $Q \equiv \sqrt{\wt} q$ which is analogous to the redefinition~\eqref{eq:FreeModelActions} of fields. In field theory, the time variable $t$ becomes the space-time variable $x$ and the function $\wt (t) \equiv \re^{\gamma t} $ becomes $ w(x) \equiv 1/\sqrt{ \det \big( \theta^{\m \n} (x) \big) }$. This analogy is of interest in view of the fact that the damped harmonic oscillator is known to possess conserved charges involving an explicit time-dependence, see Appendix~\ref{sec:dampedosci}. The generalization of the latter result to the energy-momentum tensor in field theory will be discussed in Section~\ref{sec:ScalFieldCoupled}.

\paragraph{Symmetries:} If the field $\theta^{\m \n}$ transforms under Poincar\'e transformations $x^\m \mapsto x^{\prime \m} = {\Lambda^\m}_\n x^\n + a^\m$ as a classical relativistic field, i.e., as
\begin{gather}\label{eq:PoincTrans}
\theta^{\prime \m \n} (x') = {\Lambda^\m}_\rho {\Lambda^\n}_\sigma \theta^{\rho \sigma} (x) ,
\end{gather}
then the weight function $w = 1/\sqrt{\det (\theta^{\m \n})}$ is Poincar\'e invariant and the derivative $\cd_\m$ transforms covariantly. The given action functionals are then Poincar\'e invariant (as noted in~\cite{Kupriyanov:2013jka} for the case of a Dirac field).

The action~\eqref{eq:ActScalar} for a scalar field of charge $e$ is invariant under global $U (1)$-gauge transformations whose infinitesimal form is given by
\begin{gather}\label{eq:InfGlobTrans}
{\delta} \phi = \ri e \varepsilon \phi , \qquad {\delta} \phi^* = -\ri e \varepsilon \phi ^* ,
\end{gather}
where $\varepsilon$ is a constant real parameter. For the Dirac field action~\eqref{eq:ActDirac}, we also have such an $U (1)$-invariance for the fields $\psi$, $\bar{\psi}$. By virtue of Noether's first theorem, one thus expects the existence of locally conserved current densities associated to the global $U (1)$-invariance of the models describing the scalar and Dirac fields: the corresponding expressions will be derived in Section~\ref{sec:ConsLaws}, see equations~\eqref{eq:ScalCurrDens} and~\eqref{eq:DiracCurrDens} below.

Due to the coupling of fields to the $x$-dependent external fields $\theta^{\m \n}$, one does not expect the energy-momentum tensor of matter or gauge fields to be locally conserved, even for the case of free scalar or Dirac fields. However, the analogy of the corresponding free field models with the damped harmonic oscillator in mechanics and the existence of a conserved charge for the latter dynamical system indicate that a local conservation law also holds for these models on noncommutative space-time. The corresponding expression will be derived in Section~\ref{sec:ConsLaws} along with the energy-momentum balance equation which holds for interacting fields.

\section{Simple derivation of conservation laws}\label{sec:conslaws}

Before deriving conservation laws for the field theoretical models on noncommutative space-time discussed above, we outline a simple derivation of conservation laws in Minkowski space which can be generalized straightforwardly to noncommutative space. We proceed as in non-relativistic mechanics (where the conservation of energy is obtained by multiplying the equation of motion by ${\rm d}{\vec x}/{\rm d}t$) or in nonrelativistic quantum mechanics (where the continuity equation for the probability current density is obtained by multiplying the wave equation for $\psi$ by $\psi ^*$ and then subtracting the complex conjugate expression). The application of this procedure in relativistic field theory amounts to a simple derivation of Noether's first theorem in this setting.

\subsection{General procedure}

\paragraph{General procedure in relativistic field theory:} Consider a collection $\varphi \equiv (\vp_r)_{r = 1, \dots, N}$ of classical relativistic fields in Minkowski space and suppose that their dynamics is described by an action functional $S[\varphi ] \equiv \int {\rm d}^nx {\cal L} (\varphi , \pa_{\mu} \varphi, x)$ which involves a Lagrangian density ${\cal L}$ which may explicitly depend on the space-time coordinates~$x$. The associated equations of motion are given by $\delta S /\delta \vp =0$ where the functional derivative has the following form if the Lagrangian depends at most on the first order partial derivatives of $\vp$:
 \begin{gather}\label{eq:fctderAction}
\frac{\delta S}{\delta \vp} = \frac{\pa {\cal L}}{\pa \vp} - \pa_\m \left( \frac{\pa {\cal L}}{\pa \pa_\m \vp} \right) .
\end{gather}
Let us now consider the case of an $x$-independent Lagrangian density and some \emph{active symmetry transformations} $\delta \varphi (x) \equiv \varphi ' (x) - \varphi (x) $ which depend continuously on one, or several, real constant symmetry parameters. We suppose that the Lagrangian density is \emph{quasi-invariant} under these transformations (\emph{``divergence symmetry''}), i.e., $\delta {\cal L} = \pa_\m \Omega ^\m $ for some (possibly vanishing) vector field $(\Omega^\mu)$ depending on $\vp$. Let us multiply the functional derivative~\eqref{eq:fctderAction} by $\delta \vp$ and apply the Leibniz rule to the partial derivative term:
\begin{gather}\label{eq:MultPhidot}
- \frac{\delta S}{\delta \vp} \delta \vp= \pa_\m \left( \frac{\pa {\cal L}}{\pa \pa_\m \vp} \right) \delta \vp - \frac{\pa {\cal L}}{\pa \vp} \delta \vp
 = \pa_\m \left( \frac{\pa {\cal L}}{\pa \pa_\m \vp} \delta \vp \right) - \bigg[ \underbrace{\frac{\pa {\cal L}}{\pa \pa_\m \vp} \pa_\m \delta \vp
+ \frac{\pa {\cal L}}{\pa \vp} \delta \vp }_{= \delta {\cal L} = \pa_\m \Omega ^\m }\bigg] .
\end{gather}
Thus we have
\begin{gather}
0 = \frac{\delta S}{\delta \vp} \delta \vp + \pa_\m j^\m
\qquad \mbox{with} \qquad
j^\m \equiv \frac{\pa {\cal L}}{\pa (\pa _\m \vp )} {\delta} \vp- \Omega^\m ,\label{eq:Noether1}
\end{gather}
i.e., the general form of \emph{Noether's first theorem} in relativistic field theory. This suggests a~similar procedure to be followed in the next section for noncommutative space: we multiply the functional derivative by an appropriate variation of fields and then express the product as a~total derivative $\pa_\m j^\m$, hence this derivative vanishes for the solutions of the equations of motion.

\paragraph{Some examples:} For instance, for \emph{space-time translations} (by $a \in \br^n$), we have $\delta \vp = a_\n \pa^\nu \vp$ and
$\delta {\cal L} = a_\n \pa^\nu {\cal L} = \pa_\m (\eta^{\m \n} a_\n {\cal L})$.
After factorizing the symmetry parameters,
$j^\m = a_\n T^{\m \n}_{\textrm{can}}$, we conclude that
 the local conservation law for the canonical energy-momentum tensor (EMT) $T_{\textrm{can}} ^{\m \n} $ holds
for all solutions of the equations of motion:
\begin{gather*}
 \pa_\m T_{\textrm{can}} ^{\m \n} = 0 , \qquad \mbox{with}\qquad T_{\textrm{can}} ^{\m \n} \equiv \frac{\pa {\cal L}}{\pa (\pa_{\mu} \vp)} \pa^\n \vp -\eta^{\m \n} {\cal L} .
\end{gather*}
If the Lagrangian density ${\cal L}$ depends explicitly on the space-time coordinates $x$, then the last term in equation~\eqref{eq:MultPhidot} reads $a_\n ( \pa^\n {\cal L} - \pa^\n _{\textrm{expl}}{\cal L})$ so that
 we obtain the \emph{energy-momentum balance equation}
\begin{gather}
\pa_\m T^{\m \n}_{\textrm{can}} = - \pa_{\textrm{expl}} ^\n {\cal L} ,\label{eq:balCanEMT}
\end{gather}
where $\pa_{\textrm{expl}} ^\n {\cal L}$ reflects the explicit $x$-dependence of ${\cal L}$. A simple illustration which is relevant for the coupling of matter fields to a given symplectic structure (that we address in the next subsection) is given by the linear coupling of a real scalar field~$\phi$ to a fixed $x$-dependent external source $J$, i.e., the Lagrangian density ${\cal L} = \frac{1}{2} \pa^\m \phi \pa_\m \phi - \frac{m^2}{2} \phi ^2 - J\phi$. The energy-momentum balance equation~\eqref{eq:balCanEMT} then reads
\begin{gather}
\pa_\m T^{\m \n}_{\textrm{can}} = (\pa^\n J) \phi , \qquad \mbox{with} \qquad T^{\m \n}_{\textrm{can}} = \pa^\m \phi \pa^\n \phi - \eta^{\m \n} {\cal L} .
\label{eq:balEMTsource}
\end{gather}

Another illustration of the general procedure~\eqref{eq:MultPhidot}--\eqref{eq:Noether1} is given by \emph{internal symmetries} for charged fields, e.g., for $\vp = (\phi, \phi^*)$ where $\phi$ represents a complex scalar field. The real-valued Lagrangian density is invariant under internal symmetry transformations labeled by a constant real parameter $\varepsilon$ (and the electric charge~$e$ of the field $\phi$),
\begin{gather*}
 {\delta} \phi = \ri e \varepsilon \phi , \qquad {\delta} \phi^* = -\ri e \varepsilon \phi ^* .
\end{gather*}
In this case, the procedure~\eqref{eq:MultPhidot}--\eqref{eq:Noether1} yields the off-shell identity
\begin{gather}\label{eq:ConsElCharge}
0 = \left[ \frac{\delta S}{\delta \phi} \phi - \mbox{c.c.}\right] + \pa_\m j^\m , \qquad \mbox{with}\qquad j^{\m} \equiv \frac{\pa {\cal L}}{\pa (\pa_{\mu} \phi)} \phi - \mbox{c.c.} .
\end{gather}
The local conservation law $\pa_\m j^\m =0$ (which holds for any solution of the equation of motion) now expresses the conservation of electric charge.

In gauge field theories, multiplication of the functional derivative $\delta S / \delta A^\m$ with the field strength tensor $F^{\m \n}$ readily yields the physical, gauge invariant EMT $T^{\m \n}$ of the gauge field~$(A^\m)$ whereas the standard Noether procedure leads to the canonical, non-gauge invariant EMT~$T^{\m \n} _{\textrm{can}}$ (which has to be improved by the Belinfante or other procedures~\cite{Blaschke:2016ohs}).

\subsection{Scalar field coupled to a symplectic structure}\label{sec:ScalFieldCoupled}

It is instructive to apply the general procedure outlined above to the action for a self-interacting complex scalar field coupled to a given symplectic structure $(\omega_{\m \n})$, this action being given by
expression~\eqref{eq:ActScalar} without the star products:
\begin{gather}\label{eq:ActScalSymp}
S_{\omega} [\phi,\phi^*] \equiv \int_{\br ^n} {\rm d}^n x w \left[ \cd^\rho \phi^* \cd_\rho \phi - m^2 \phi^* \phi - \frac{\cg}{2} (\phi^* \phi)^2\right] \equiv \int_{\br ^n} {\rm d}^n x w {\cal L}_{\omega} (\phi,\phi^*) .
\end{gather}
We presently multiply the functional derivatives $\delta S_{\omega} /\delta \phi$ and $\delta S_{\omega} /\delta \phi^*$ by the covariant derivatives of fields and apply the Leibniz rule for these derivatives as well as the relation $[{\cal D}_\m , {\cal D}_\n ] =0$:
\begin{gather}
- w \left[ \frac{\delta S_{\omega}}{\delta \phi} \cd^\n \phi + \cd^\n \phi^* \frac{\delta S_{\omega}}{\delta \phi^*} \right] =w \cd_\m \big[ \cd^\m \phi^* \cd^\n \phi + \cd^\n \phi^* \cd^\m \phi - \eta^{\m \n} {\cal L}_{\omega}\big] .\label{eq:StartDerEMT}
\end{gather}
For the covariant derivatives, we have to keep in mind the weight of the fields on which they act, e.g., $\phi$, $\phi^*$ have weight $1/2$, $\phi \phi^*$ has weight $1$ and $(\phi \phi^*)^2$ has weight $2$, hence $\cd_\m (\phi \phi^*)^2 = \pa_\m (\phi \phi^*)^2 +2 \pa_\m (\ln w ) (\phi \phi^*)^2$. From this fact we deduce that equation~\eqref{eq:StartDerEMT} reads as follows for the solutions of the equations of motion (for which the left hand side of~\eqref{eq:StartDerEMT} vanishes):
\begin{gather}
0= \pa_\m T^{\m \n}_{\omega} + \frac{\cg}{2} (\pa^\n w) ( \phi \phi^*)^2 \qquad \mbox{with} \qquad
 T^{\m \n}_{\omega} = w \big[ \cd^\m \phi^* \cd^\n \phi + \cd^\n \phi^* \cd^\m \phi- \eta^{\m \n} {\cal L}_{\omega} \big] .\!\!\!\label{eq:balEMTomega}
\end{gather}
This result means that the EMT $ T^{\m \n}_{\omega}$ of matter fields is conserved for $\cg =0$, i.e., in the absence of a self-interaction, though it is not in the presence of the latter. This result can be traced back to the fact that the matter fields are coupled to a fixed external field $(\omega _{\m \n})$ (by means of the variable $w = \sqrt{\det (\omega _{\m \n})}$) and is reminiscent of the coupling of a scalar field to an external scalar source described above, see~equation~\eqref{eq:balEMTsource}.

The previous conclusions can be further elucidated by rescaling the matter fields as we did in equation~\eqref{eq:FreeModelActions}: with $\Phi \equiv \sqrt{w} \phi$ and $\Phi^* \equiv \sqrt{w} \phi^*$, the action~\eqref{eq:ActScalSymp} reads
\begin{gather}
S_{\omega} [\Phi,\Phi^*] = \int_{\br ^n} {\rm d}^n x \left[ \pa^\rho \Phi^* \pa_\rho \Phi - m^2 \Phi^* \Phi - \frac{\cg}{2} \frac{1}{w} (\Phi^* \Phi)^2\right]\equiv \int_{\br ^n} {\rm d}^n x {\cal L}_{\omega} (\Phi,\Phi^*) ,
\label{eq:InteractActions}
\end{gather}
i.e., the external field $w$ now only appears in the last term of the Lagrangian\footnote{We will come back to the factor $1/w$ (and the introduction of more general functions of $w$) at the end of~Section~\ref{sec:CurvedST}.}. In terms of $\Phi$ and $\Phi^*$, the EMT~\eqref{eq:balEMTomega} reads
\begin{gather*}
 T^{\m \n}_{\omega} = \pa^\m \Phi^* \pa^\n \Phi + \pa^\n \Phi^* \pa^\m \Phi- \eta^{\m \n} {\cal L}_{\omega} ,
\end{gather*}
and one can readily verify the energy-momentum balance equation~\eqref{eq:balEMTomega} by using the equation of motion $0= \big(\Box +m^2\big)\Phi + \cg \frac{1}{w} \Phi^* \Phi ^2$ and its complex conjugate. For $\cg =0$, we have the analogy with the damped harmonic oscillator in classical mechanics pointed out in equa\-tions~\eqref{eq:FreeModelActions}--\eqref{eq:EOMdampOsc}. The expression~\eqref{eq:ConsCharge} of the conserved charge for the latter dynamical system then allows us to get a physical understanding of the local conservation law $\pa_\m T^{\m \n}_{\omega} =0$ for matter fields which holds for $c=0$ despite their coupling to an external field $w$: in the course of the temporal evolution of matter fields $\phi$, the presence of $w$ in $T^{\m \n}_{\omega}$ is compensated by the dependence on $w$ of the solutions $\phi$ of field equations, thus ensuing the existence of conserved charges $\int_{\br ^{n-1}} {\rm d}^{n-1} x T^{0 \n}_{\omega}$ for $\n \in \{ 0, 1, \dots, n-1 \}$.

\section{Conservation laws for noncommutative field theories}\label{sec:ConsLaws}

In the previous section, we presented a derivation of the local conservation laws for the charge and for the energy-momentum which only relies on the equations of motion. The application of this procedure for Lagrangian field theories on noncommutative space simply consists of replacing the multiplication by a field $\vp$ (or by its derivative $\pa^\n \vp$) by the star product with this field (or by its covariant derivative $\cd^\n \vp$). Furthermore, for the differentiation of star products, one has to consider the product rule~\eqref{eq:IdSP} which applies to this kind of products. (We note that Noether's first theorem for \emph{constant} noncommutativity parameters has been discussed by numerous authors, e.g., see~\cite{Balasin:2015hna, Geloun:2007zza,Schweda:2000,Micu:2000, Pengpan:2000kd} and references therein.)

\subsection{Charge conservation law}

For the \emph{complex scalar field} described by the action~\eqref{eq:ActScalar} and the associated functional derivatives~\eqref{eq:FunctDerScal}, the procedure~\eqref{eq:ConsElCharge} (generalized to the {\nc} setting~\cite{Kupriyanov:2015uxa}) yields upon application of the product rule~\eqref{eq:IdSP}
\begin{gather*}
- \ri e w \left[ \phi^* \star \frac{\delta S}{\delta \phi^*} - \frac{\delta S}{\delta \phi} \star \phi \right] = \pa_\m j^\m [\phi] ,
\end{gather*}
with
\begin{gather} \label{eq:ScalCurrDens}
 j^\m [\phi] = \ri e w \big[ \phi^* \star \cd^\m \phi - \cd^\m \phi^* \star \phi \big] + \ri e \big[ b^\m _\n (\phi^*, \cd^\n \phi ) - b^\m _\n (\cd^\n \phi^* , \phi ) \big] .
\end{gather}
We note that by symmetrizing these expressions with respect to $\phi$ and $\phi^*$, we can eliminate the terms of $b^\m_\n$ which are linear in $\theta$ since $b^\m_\n (f,g)$ is antisymmetric in $f$,~$g$ at this order:
\begin{gather*}
- \frac{\ri}2 e w\left[ \staraco{\phi^*}{\var{S}{\phi^*}} - \staraco{\var{S}{\phi}}{\phi} \right] = \pa_\m \hat j^\m [\phi] ,
\end{gather*}
with
\begin{gather*}
 \hat j^\m [\phi] = \frac{\ri}2 e w \big[ \staraco{\phi^*}{\cd^\m \phi} - \staraco{\cd^\m \phi^*}{\phi} \big] +\cO\big(\th^2\big) .
\end{gather*}
In the particular case where $n=3$ and where the noncommutativity is given by $\theta ^{ij} = \varepsilon^{ijk} x^k$ with $i,j,k \in \{ 1,2,3 \}$ (in which case one can choose $w \equiv 1$ \cite{Kupriyanov:2015uxa}), the result~\eqref{eq:ScalCurrDens} coincides, upon exchange $\phi \leftrightarrow \phi^*$, with the one derived in~\cite{Kupriyanov:2015uxa}.

For the \emph{Dirac field} described by the action \eqref{eq:ActDirac} and the associated functional derivatives \eqref{eq:FunctDerDir}, the same procedure gives (in agreement with~\cite{Kupriyanov:2013jka})
\begin{gather}\label{eq:DiracCurrDens}
- \ri e w \left[ \bar{\psi} \star \frac{\delta S}{\delta \bar{\psi}} - \frac{\delta S}{\delta \psi} \star \psi \right] = \pa_\m j^\m [\psi] , \qquad\! \mbox{with} \qquad\!
 j^\m [\psi] = ew ( \bar{\psi} \gamma^\m \star \psi ) +e b^\m _\n (\bar{\psi} \gamma^\n , \psi ) .\!\!\!\!
\end{gather}
Once more, the terms in $b^\m_\n$ which are of order $\theta$ can be eliminated by symmetrization.

These conservation laws (which hold for all solutions of the equations of motion) reflect the invariance of the underlying models under the global $U (1)$-gauge transformations~\eqref{eq:InfGlobTrans}.

The conserved electric charge $Q$ then has the standard form $\int_{\br ^{n-1}} {\rm d}^{n-1} x j^0$, e.g., for the Dirac field,
\begin{gather*}
Q = \int_{\br ^{n-1}} {\rm d}^{n-1} x \big[ ew {\psi}^{\dagger} \psi +e b^0 _\n (\bar{\psi} \gamma^\n , \psi ) \big] .
\end{gather*}
In the commutative limit, the conserved current densities reduce to the familiar expressions $j^\m [\phi]= \ri e [ \phi^* \pa^\m \phi - \phi \pa^\m \phi^* ]$ and $j^\m [\psi] = e \bar{\psi} \gamma^\m \psi$.

\subsection{Energy-momentum conservation law}

\paragraph{Free field case:} We start with the case of a \emph{free complex scalar field} $\phi$ described by the action~\eqref{eq:ActScalar} with $c=0$, and by the associated functional derivatives~\eqref{eq:FunctDerScal}. By following the procedure~\eqref{eq:MultPhidot}--\eqref{eq:Noether1} for the collection of fields $\vp = (\phi , \phi^*)$, we have
\begin{gather}
- w \left[ \frac{\delta S}{\delta \phi} \star \cd^\n \phi +\cd^\n \phi^* \star \frac{\delta S}{\delta \phi^*} \right] \nonumber\\
\qquad{} =w \big[ \cd _\m \cd^\m \phi ^* \star \cd^\n \phi + \cd^\n \phi^* \star \cd _\m \cd^\m \phi
+ m^2 \big( \phi ^* \star \cd^\n \phi + \cd^\n \phi^* \star \phi \big)\big] .\label{eq:Start}
\end{gather}
By virtue of the product rule~\eqref{eq:IdSP}, the term proportional to $m^2$ reads
\begin{gather*}
\eta^{\n \m} m^2\big\{ \pa_\m \big[ w ( \phi^* \star \phi ) \big] + \pa_\rho b^\rho_\m ( \phi^* , \phi )\big\} .
\end{gather*}
For the second order derivative terms, we add and subtract the terms which are missing in order to apply relation~\eqref{eq:IdSP} with respect to the indices $\m$ and $\n$:
\begin{gather*}
w \big[ \cd _\m \cd^\m \phi ^* \star \cd^\n \phi +\cd _\m \phi ^* \star \cd^\m \cd^\n \phi\big] + w \big[ \cd _\m \cd^\n \phi^* \star \cd^\m \phi+ \cd^\n \phi^* \star \cd _\m \cd^\m \phi\big]\\
\qquad {} - w \big[ \cd _\m \phi ^* \star \cd^\m \cd^\n \phi+ \cd _\m \cd^\n \phi^* \star \cd^\m \phi\big] .
\end{gather*}
For the second order derivatives in the last line, we use the fact that the covariant derivatives commute with each other. Thus, relation~\eqref{eq:IdSP} allows us to rewrite each of the three expressions appearing in the previous equation as a total derivative. Altogether we obtain the off-shell identity
\begin{gather}\label{eq:RelEMTScalField}
- w \left[ \frac{\delta S}{\delta \phi} \star \cd^\n \phi+\cd^\n \phi^* \star \frac{\delta S}{\delta \phi^*} \right] = \pa_\m T^{\m \n}_{\textrm{free}} [\phi] ,
\end{gather}
involving the (on-shell conserved) \emph{EMT of the scalar field} given by
\begin{gather}
 T^{\m \n}_{\textrm{free}} [\phi] =
w \big[ \big( \cd ^\m \phi ^* \star \cd^\n \phi + \cd ^\n \phi ^* \star \cd^\m \phi \big)
- \eta^{\m \n} \big( \cd ^\rho \phi ^* \star \cd_\rho \phi - m^2 \phi ^* \star \phi \big) \big]\nonumber\\
\hphantom{T^{\m \n}_{\textrm{free}} [\phi] =}{} + b^\m _\rho ( \cd ^\rho \phi ^* , \cd^\n \phi )+ b^\m _\rho ( \cd ^\n \phi ^* , \cd^\rho \phi )- \big[ b^{\m \n} ( \cd ^\rho \phi ^* , \cd_\rho \phi )
-m^2 b^{\m \n} ( \phi ^* , \phi )\big] ,\label{eq:EMTNCscal}
\end{gather}
where $b^{\m \n} \equiv \eta^{\n \sigma} b^\m _\sigma$. For the case of constant noncommutativity parameters, this tensor reduces (up to a multiplicative constant) to the expression
\begin{gather*}
 T_{\textrm{free,} \theta = \textrm{const}} ^{\m \n} [\phi] = \pa ^\m \phi ^* \star \pa^\n \phi + \pa ^\n \phi ^* \star \pa^\m \phi- \eta^{\m \n} \big( \pa ^\rho \phi ^* \star \pa_\rho \phi - m^2 \phi ^* \star \phi \big) .
\end{gather*}
For real-valued fields $\phi$ we thus recover the well-known result which has been obtained by other arguments in the literature~\cite{Micu:2000}. In the commutative limit, the expression $ T_{\textrm{free,} \theta = \textrm{const}}$ reduces to the familiar result from Minkowski space~\cite{Blaschke:2016ohs}.

As for the derivation of the local conservation law of electric charge, one can start from a~symmetrized expression in~equation~\eqref{eq:Start}, i.e., replace the star products by star anticommutators: this again allows us to eliminate in the final result the terms of $b^\m_\n$ which are linear in $\theta$.

For the \emph{free Dirac field,} the same line of reasoning (see~\cite{Kupriyanov:2013jka} for the case of quantum mechanics) yields an equation which is completely analogous to equation~\eqref{eq:RelEMTScalField}:
\begin{gather*}
- w \left[ \frac{\delta S}{\delta \psi} \star \cd^\n \psi + \cd^\n \bar{\psi} \star \frac{\delta S}{\delta \bar{\psi}} \right]= \pa_\m T^{\m \n}_{\textrm{free}} [\psi] .
\end{gather*}
Upon use of the equations of motion, the contributions to the \emph{EMT} $T^{\m \n}_{\textrm{free}} [\psi]$ \emph{of the Dirac field} which involve $\eta^{\m \n}$ and $ \eta^{\n \sigma} b^\m _\sigma$ vanish and we are left with the expression
\begin{gather}
 T^{\m \n}_{\textrm{free}} [\psi] =w \tfrac{\ri}{2} \big( \bar{\psi} \gamma^\m \star \cd^\n \psi -\cd ^\n \bar{\psi} \star \gamma^\m \psi \big)
+ b^\m _\rho \big( \bar{\psi} , \tfrac{\ri}{2} \gamma^\rho \cd^\n \psi \big) - b^\m _\rho \big( \tfrac{\ri}{2} \cd^\n \bar{\psi} \gamma^\rho , \psi \big) .\label{eq:EMTNCDirac}
\end{gather}
In the commutative limit, we again recover the familiar expression $T^{\m \n}_{\textrm{free},\theta =0} [\psi] =\frac{\ri}{2} \big(\bar{\psi} \gamma^\m \pa^\n \psi -\pa ^\n \bar{\psi} \gamma^\m \psi \big)$.

The fact that the EMT of matter fields is locally conserved despite the coupling of these fields to the space-time dependent external field $\theta^{\m \n}(x)$ is somewhat unexpected. However, in this respect it is worthwhile to remember the close analogies which exist in various instances between the noncommutativity parameters and a magnetic field, as well as the fact that the energy of a charged particle coupled to a magnetic field is conserved. Furthermore, the analogy that we described in equations~\eqref{eq:FreeModelActions}--\eqref{eq:EOMdampOsc} between the dynamics of free matter fields in noncommutative space-time and the damped harmonic oscillator in classical mechanics allows us to get an understanding for the appearance and form of conserved quantities: For the damped harmonic oscillator, the conserved quantity~\eqref{eq:ConsCharge} does not represent the total energy $H_0$ of the (undamped) oscillator (which is not conserved) and analogously expressions~\eqref{eq:EMTNCscal} and~\eqref{eq:EMTNCDirac} represent locally conserved quantities for fields which are coupled to $(\theta^{\m \n})$ rather than the EMT of uncoupled fields.

The symmetry which is at the origin of the local conservation law~\eqref{eq:EMTNCscal} can easily be identified in view of the derivation~\eqref{eq:RelEMTScalField} of this conservation law (and similarly for the Dirac field). Indeed, by using the fact that the covariant derivatives $\cd_\m$ and $\cd_\rho$ commute with each other and by applying the modified Leibniz rule~\eqref{eq:LeibnizRule}, one can check that the free scalar field action (i.e., expression~\eqref{eq:ActScalar} with $c=0$) is invariant under the infinitesimal transformations (parametrized by $a^\m \in \br$)
 \begin{gather}
\delta \phi = a^\m \cd_\m \phi , \qquad \delta \phi ^* = a^\m \cd_\m \phi ^* ,\label{eq:infSymTransf}
\end{gather}
due to the fact that the integrand of the action transforms into a total derivative. Thus, we have a~divergence symmetry parametrized by the real constants~$a^\m$. In fact, the variation $\delta _\m \phi \equiv \cd_\m \phi = \big[ \pa_\m + \big(\pa_\m \ln w ^{1/2} \big) \big] \phi$ is completely analogous to the infinitesimal symmetry transformation~\eqref{eq:SymTransfDHO} (i.e., $\delta q = \big[ \pa_t + \big(\pa_t \ln \wt ^{1/2} \big) \big] q$)
which is at the origin of the conserved quantity~\eqref{eq:ConsCharge} for the damped harmonic oscillator. Although the variation $\delta \phi = a^\m \cd_\m \phi = a^\m \pa_\m \phi + c\phi $ with $c \equiv \pa_\m \big(a^\m \ln w ^{1/2} \big)$ involves the infinitesimal translation $ a^\m \pa_\m \phi$ of the field $\phi$, the symmetry transformations~\eqref{eq:infSymTransf} cannot be identified with the Poincar\'e invariance of the action that we pointed out at the end of Section~\ref{sec:FTNC} since this invariance assumes that the field $(\theta ^{\m \n})$ transforms as a classical relativistic field under Poincar\'e transformations: the latter transformation of~$\theta^{\m \n}$ is ``compensated'' here by the incorporation of the local, $\theta$-dependent scale transformation $\delta _c \phi \equiv c\phi = \pa_\m \big(a^\m \ln w ^{1/2} \big) \phi$ which ensures that the variation $ a^\m \pa_\m \phi + c\phi = \delta \phi$ represents a~symmetry transformation giving rise to the local conservation law~\eqref{eq:EMTNCscal}.

\paragraph{Case of interacting fields:} For the \emph{self-interacting complex scalar field} $\phi$, we have an additional contribution on the right hand side of equation~\eqref{eq:Start}:
\begin{gather*}
w \cg \big[ \phi ^* \star \phi \star \phi ^* \star \cd^\n \phi + \cd^\n \phi^* \star \phi \star \phi ^* \star \phi\big] .
\end{gather*}
This term can be rewritten as a sum $C^\n +B^\n$ where $C^\n$ is a sum of star-commutators,
\begin{gather}
C^\n = w \frac{\cg}{2} \big(\starco{\phi ^* \star \phi}{\phi ^* \star \cd^\n \phi} + \starco{\cd^\n \phi^* \star \phi}{\phi ^* \star \phi}\big) ,\label{eq:StarComEMT}
\end{gather}
and where $B^\n$ is given by
\begin{gather*}
B^\n = w \frac{\cg}{2} \big( \cd^\n \phi^* \star \phi \star \phi ^* \star \phi+\phi^* \star \cd^\n \phi \star \phi ^* \star \phi+\phi^* \star \phi \star \cd^\n \phi ^* \star \phi+\phi ^* \star \phi \star \phi ^* \star \cd^\n \phi\big) .
\end{gather*}
A quartic star monomial has weight $2$, hence
\begin{gather*}
\pa^\n \big[ w (\phi^* \star \phi \star \phi ^* \star \phi) \big] = w \cd ^\n (\phi^* \star \phi \star \phi ^* \star \phi) - (\pa^\n w) \phi^* \star \phi \star \phi ^* \star \phi .
\end{gather*}
From this relation and from the modified Leibniz rule~\eqref{eq:LeibnizRule} it follows that
\begin{gather*}
B^\n = \pa_\m \left\{ \eta^{\m \n} \big[ w \frac{\cg}{2} ( \phi^* \star \phi) \star (\phi ^* \star \phi ) + \frac{\cg}{2} b^{\m \n} ( \phi^* \star \phi , \phi ^* \star \phi ) \big] \right\}\\
\hphantom{B^\n =}{} + \frac{\cg}{2} (\pa^\n w)\phi^* \star \phi \star \phi ^* \star \phi + A^\n ,
\end{gather*}
where $A^\n$ represents a star-anticommutator, $A^\n \equiv w \frac{\cg}{2} \staraco{\phi ^* \star \phi}{w^{-1} \pa_\m b^{\m \n} (\phi ^* , \phi) }$. Thus, the \emph{EMT for the self-interacting theory} reads
\begin{gather*}
 T^{\m \n} [\phi] \equiv T_{\textrm{free}} ^{\m \n} [\phi]
- w \eta^{\m \n} \left( -\frac{\cg}{2} \phi^* \star \phi \star \phi ^* \star \phi \right) + \frac{\cg}{2} b^{\m \n} ( \phi^* \star \phi , \phi ^* \star \phi ),
\end{gather*}
and, for the solutions of the equations of motion, we have the \emph{energy-momentum balance equation}
\begin{gather}
\pa_\m T^{\m \n} [\phi] = - \frac{\cg}{2} (\pa^\n w)\phi^* \star \phi \star \phi ^* \star \phi- (C^\n + A^\n) . \label{eq:NonConsEMT}
\end{gather}
If we integrate this relation over space-time, then the integral over $C^\n$ (i.e., a sum of star-commutators) vanishes due to the cyclicity of the trace~\eqref{eq:Closure}. However, this is not the case for the other terms on the right hand side of~\eqref{eq:NonConsEMT}. The non-conservation law of $ T^{\m \n} [\phi]$ for interacting fields is related to the fact that the matter field $\phi$ is coupled to the external tensor field $(\theta^{\m \n} (x))$ and its derivatives as noted already in Section~\ref{sec:ScalFieldCoupled} for the coupling of a scalar field to a symplectic structure.

For constant noncommutativity parameters, we have $w=\textrm{const}$ and $\cd_\m \phi = \pa_\m \phi$: the result~\eqref{eq:NonConsEMT}, with $C^\n$ given by~\eqref{eq:StarComEMT}, can then be checked readily by using the equations of motion. For real-valued fields, the latter result reduces to $\frac{\cg}{2} \starco{\phi \star \phi}{\starco{\phi}{\pa^\n \phi} }$, i.e., the result which was first obtained in~\cite{Micu:2000} by other methods (and which has been further discussed in~\cite{Schweda:2000}).

For the \emph{gauge field} $(A^\m )$, an energy-momentum balance equation can be obtained by starting from the product $-w \frac{\delta S}{\delta A_\n} F_{\n \m}$. We will not expand further on this point since it is already fairly involved in the case of constant noncommutativity parameters, see~\cite{Balasin:2015hna} and references therein.

\section{Field theory on curved noncommutative space-time}\label{sec:FTcurvedSpace}

With the description of gravity in mind, the formulation of noncommutative field theories (and in particular of gauge theories) on generic symplectic manifolds with curvature and/or torsion has been addressed by various authors using diverse approaches, e.g., see~\cite{Aschieri:2005yw,Aschieri:2009zz, Aschieri:2005zs, Bayen:1977ha, Beggs:2017yzq,Calmet:2005qm, Chaichian:2010yt, Chaichian:2010yw,Chamseddine:2003cw,Chamseddine:2010ps, Ciric:2018ygk, Dobrski:2010pu, Dobrski:2015emm, Fedosov:1996, Fritz:2016svs, GraciaBondia:2010sm, Hawkins:2002rf, Hawkins:2007, McCurdy:2009xz, Schenkel:2012zu,Szabo:2006wx, Vassilevich:2009cb, Vassilevich:2010he} as well as~\cite{Mueller-Hoissen, Schenkel:2010sc} for some nice introductions and overviews of the literature up to the year 2010. In relationship with the main subject of the present work (in particular the
conservation laws for field theories on flat noncommutative space-time) we note that it should also be possible to obtain the energy-momentum tensor (EMT) of matter fields in flat space-time by coupling these fields to a metric tensor field: the EMT is then given by the flat space limit of the curved space EMT defined as the variational derivative of the matter field action with respect to the metric tensor (see~\cite{Blaschke:2016ohs} and references therein for a justification of this procedure). Here we outline the approach to curved noncommutative space which was recently put forward by M.~Dobrski~\cite{Dobrski:2015emm} who discussed the case of pure gravity
following a series of related works by the same author, notably~\cite{Dobrski:2010pu}: this formulation appears to fit nicely with the one that we considered here for flat noncommutative space-time. In a separate work (in preparation), we further discuss star products on curved manifolds and in particular different approaches to the description of tensor fields and differential forms on noncommutative manifolds.

\subsection{Curved space-time and symplectic structure}\label{sec:CurvedST}

So far we discussed the star product on $\br ^n$ (with $n$ even) where $\br ^n$ is considered as a flat symplectic manifold, i.e., as a flat smooth manifold equipped with a symplectic two-form $\underline{\omega} \equiv \frac{1}{2} \omega_{\m \n} {\rm d}x^\m \wedge {\rm d}x^\n$. In Einstein's theory of gravity, the gravitational field is described by an $x$-dependent metric, i.e., the space-time manifold $M$ is endowed with a symmetric tensor field $\underline{g} \equiv g_{\m \n} {\rm d}x^\m \otimes {\rm d}x^\n$. Whatever the manifold under consideration, the definition of a parallel transport of vectors (and more generally of tensor fields), requires the introduction of a \emph{linear connection}~$\nabla$: Its action on tensor fields $(V^\m )$ or $(V_\m)$ is locally defined in terms of the \emph{connection coefficients} $\Gamma^\lambda _{\m \n}$ of $\nabla$, namely~\cite{straumann} (with the notation $\nabla_{\pa_\m} \equiv \nabla_\m$)
\begin{gather*}
\nabla_\m V^\lambda \equiv \pa_\m V^\lambda + \Gamma^\lambda _{\m \n} V^\n , \qquad \nabla_\m V_\lambda \equiv \pa_\m V_\lambda - \Gamma^\rho _{\m \lambda} V_\rho .
\end{gather*}
To this connection one associates its \emph{curvature} and its \emph{torsion} given by the tensor fields $\big(R^\rho _{\ \sigma \m \n}\big)$ and $\big(T^\lambda _{\m \n}\big) $ defined by the relation
\begin{gather*}
[ \nabla_\m , \nabla_\n ] V^\lambda = R^\lambda _{\ \sigma \m \n} V^\sigma- T^\sigma _{\m \n} \nabla_\sigma V^\lambda .
\end{gather*}
Linear connections exist on any smooth manifold $M$ and if no further assumption is made they are independent of other structures on $M$ like the metric structure or the symplectic structure. In the following, we successively consider the cases where $M$ is endowed with a metric structure, with a symplectic structure and with both structures.

\paragraph{Metric structure:} We recall that on a pseudo-Riemannian manifold $(M, \underline{g})$, there exists a~unique linear connection $\nabla$ (referred to as the \emph{Levi-Civita connection}, its connection coefficients being referred to as the Christoffel symbols) which is characterized by the following two pro\-per\-ties:
\begin{subequations}\label{eq:LeviCivita1-*}
\begin{alignat}{3}
& \nabla = \text{torsionless:} \quad && 0 = T^\lambda _{\m \n} = \Gamma^\lambda _{\m \n} - \Gamma^\lambda _{\n \m} ,& \label{eq:LeviCivita1}\\
& \nabla = \text{metric compatible:}\quad && 0 = \nabla_\lambda g_{\m \n} .& \label{eq:LeviCivita2}
\end{alignat}
\end{subequations}
Thus, the connection coefficients $ \Gamma^\lambda _{\m \n}$ are symmetric in the indices $\m$, $\n$ and the metric is covariantly constant with respect to $\nabla$. These relations imply the well-known expression for the Christoffel symbols, i.e., $\Gamma^\lambda _{\m \n} = \frac{1}{2} g^{\lambda \rho} (\pa_\m g_{\rho \n} + \pa_\n g_{\m \rho} - \pa_\rho g_{\m \n})$ which implies $ \Gamma^\n _{\m \n} = \pa_\m \textrm{ln} \sqrt{|g|}$ where $g \equiv \det (g_{\m \n})$. This connection is used in Einstein's theory of gravity and we will also consider it here for the pseudo-Riemannian manifold $(M, \underline{g} )$ while denoting it as above by $\nabla$ with the connection coefficients $\Gamma^\lambda _{\m \n}$.

\paragraph{Symplectic structure:} Since we want the generalize our description of noncommutative field theory in flat space to a more general manifold $M$, we suppose that the latter manifold is endowed with a~symplectic two-form $\underline{\omega} \equiv \frac{1}{2} \omega_{\m \n} {\rm d}x^\m \wedge {\rm d}x^\n$. Like the metric $(g_{\m \n})$, the symplectic tensor~$(\omega_{\m \n})$ is given by a non-degenerate matrix, and it is thus natural to consider a linear connection (which we denote by $\stackrel{\circ}{\nabla}$ with connection coefficients ${\stackrel{\smash{\circ}}{\Gamma}} ^\lambda _{\m \n}$, curvature ${\stackrel{\smash{\circ}}{R}} ^\rho _{\ \sigma \m \n}$ and torsion~${\stackrel{\smash{\circ}}{T}} ^\lambda _{\m \n}$), which has properties that are completely analogous to~\eqref{eq:LeviCivita1-*}: Indeed, on any symplectic
manifold it is possible~\cite{Bieliavsky, Gelfand, Waldmann:2007} to find a linear connection $\stackrel{\circ}{\nabla}$ with the properties
\begin{subequations}
\begin{alignat}{3}
& \stackrel{\circ}{\nabla} = \text{torsionless:} \quad && 0 = {\stackrel{\smash{\circ}}{T}} ^\lambda _{\m \n} = {\stackrel{\smash{\circ}}{\Gamma}} ^\lambda _{\m \n} - {\stackrel{\smash{\circ}}{\Gamma}} ^\lambda _{\n \m},& \label{eq:SympConn1}\\
& \stackrel{\circ}{\nabla} = \text{symplectic:} \quad && 0 = \stackrel{\circ}{\nabla} _\lambda \omega_{\m \n} .\label{eq:SympConn2}
\end{alignat}
\end{subequations}
We note that in the literature~\cite{Bieliavsky, Gelfand, Waldmann:2007} a `symplectic connection' is generally required to be torsionless, but we do not include this condition in our definition of `symplectic' (hence we should rather use the terminology `almost symplectic'~\cite{Bieliavsky, McCurdy:2009xz}). A~symplectic manifold $(M , \underline{\omega})$ equipped with a torsionless, symplectic connection is referred to as a \emph{Fedosov manifold}, e.g., see~\cite{Bieliavsky,Gelfand} for a general study. We remark that the antisymmetry of $(\omega_{\m \n})$ and the relation $\stackrel{\circ}{\nabla} _\lambda \omega_{\m \n} =0$ imply the closedness relation ${\rm d}\underline{\omega} =0$ since $\stackrel{\circ}{\nabla} _\lambda \omega_{\m \n} + \mbox{cyclic permutations of} (\lambda, \m , \n) = \pa_\lambda \omega_{\m \n} + \mbox{cyclic permutations of} (\lambda, \m , \n) $.

\paragraph{Metric/symplectic compatible structure:} For the formulation of gravity on the even-dimensional space-time manifold $(M , \underline{g}, \underline{\omega} )$, it is natural to relate $(\omega_{\m \n})$ and $ (g_{\m \n})$, or at least to ensure their compatibility for the parallel transport of vectors. Different conditions or relations for the connection coefficients ${\stackrel{\smash{\circ}}{\Gamma}} ^\lambda _{\m \n}$
and ${\Gamma} ^\lambda _{\m \n}$ can be envisaged~\cite{Dobrski:2015emm, Steinacker:2012ct}. The strongest condition which consists of equating both connections is very stringent since this condition entails
that~$\underline{\omega}$ is covariantly constant with respect to the Levi-Civita connection~$\nabla$: in four dimensions this implies that the metric locally decomposes into a sum of two-dimensional metrics~\cite{Dobrski:2015emm,Stephani:2003tm}. A~weaker compatibility condition was considered by M.~Dobrski who referred to it as a~\emph{weakly compatible metric/Fedosov structure}:
\begin{gather}
{\Gamma} ^\n _{\m \n} = {\stackrel{\smash{\circ}}{\Gamma}} ^\n _{\m \n} .\label{eq:WCM}
\end{gather}
Since the symplectic condition~\eqref{eq:SympConn2} yields
\begin{gather*}
{\stackrel{\smash{\circ}}{\Gamma}} ^\n _{\m \n} = \frac{1}{2} \theta^{\rho \sigma} \pa_\m \omega_{\sigma \rho} =
\pa _\m \ln w \qquad \mbox{with} \qquad w = \sqrt{\det (\omega _{\m \n})} ,
\end{gather*}
relation~\eqref{eq:WCM} is equivalent to
\begin{gather}\label{eq:GpropW}
\pa_\m \ln \sqrt{|g|} = \pa _\m \ln w , \qquad \mbox{i.e.,} \qquad \sqrt{|g|} = a w ,
\end{gather}
with some strictly positive constant $a$. This means that the Riemann integration measure and the Liouville (symplectic) integration measure are proportional to each other (with a coefficient of proportionality which may be different for different connected components of the manifold~$M$). Moreover, the covariant divergences of an antisymmetric tensor field like the Poisson tensor field~$(\theta ^{\m \n})$ with respect to the connections $\nabla$ and $\stackrel{\circ}{\nabla}$ now coincide with each other~\cite{Dobrski:2015emm} since we have
\begin{gather}
\nabla_\m \theta ^{\m \n} = \frac{1}{ \sqrt{|g|}} \pa_\m \big( \sqrt{|g|} \theta ^{\m \n} \big)
= \frac{1}{w} \pa_\m \big( w \theta ^{\m \n} \big) = \stackrel{\circ}{\nabla} _\m \! \theta ^{\m \n} .\label{eq:CovDiv}
\end{gather}
Thus, the divergenceless condition $\pa_\m \big( w \theta ^{\m \n} \big) =0$ that we imposed for $M=\br^n$ in equation~\eqref{eq:PDEw} in order to have a closed star product amounts, on a general manifold $M$, to the vanishing of expression~\eqref{eq:CovDiv}.

\paragraph{Dynamics of fields for a weakly compatible metric/Fedosov structure:} Let us now consider a space-time manifold $(M , \underline{g}, \underline{\omega} )$ of dimension $n=2m$ together with a weakly compatible metric/Fedosov structure, i.e., connections $\nabla$, $\stackrel{\circ}{\nabla}$ satisfying the set of relations~\eqref{eq:LeviCivita1-*}--\eqref{eq:WCM}. The dynamics of matter fields (e.g., of a scalar field $\phi$) which are coupled to the gravitational field described by the metric tensor field $(g_{\m \n})$ can presently be described~\cite{Dobrski:2015emm} by the standard action functionals though involving the integration measure
${\underline{\omega}^m}/{m!}$ (see equation~\eqref{eq:VolElem}). Thus, the variables $\omega_{\m \n}$ become \emph{dynamical fields} which only couple to the gravitational fields $g_{\m \n}$ by means of their respective determinants. The set of anholonomic constraints $\stackrel{\circ}{\nabla} _\lambda \omega_{\m \n} =0$, ${\stackrel{\smash{\circ}}{\Gamma}} ^\lambda _{[ \m \n ]} =0$ (where the square brackets denote antisymmetrization, as usual) and ${\stackrel{\smash{\circ}}{\Gamma}} ^\n _{\m \n} = {\Gamma} ^\n _{\m \n} $ can be imposed by a set of Lagrange multiplier tensor fields $m^{\lambda \m \n} = - m^{\lambda \n \m}$, $t^{\m \n} _\lambda = - t^{\n \m} _\lambda $ and~$s^\m$. Thus, we have a \emph{generally covariant total action} (with $\kappa \equiv 8 \pi G$ where $G$ denotes Newton's constant) given by
\begin{gather}\label{eq:actionMS1}
S \equiv S_g + S_{M} + S_{L} , \qquad \mbox{with} \qquad S_g \equiv \frac{1}{2 \kappa} \int_{M} \frac{\underline{\omega}^m}{m!} R , \qquad S_{M} \equiv \int_{M} \frac{\underline{\omega}^m}{m!} {\cal L}_{M} (\phi, g_{\m \n}) ,
\end{gather}
and
\begin{gather}\label{eq:actionMS2}
S_{L} \equiv \int_{M} \frac{\underline{\omega}^m}{m!}\left[ m^{\lambda \m \n} \stackrel{\circ}{\nabla} _\lambda \omega_{\m \n}
+ t^{\m \n} _\lambda {\stackrel{\smash{\circ}}{\Gamma}} ^\lambda _{[ \m \n ]}+ s^\m \big( {\Gamma} ^\n _{\m \n} - {\stackrel{\smash{\circ}}{\Gamma}} ^\n _{\m \n} \big)\right] .
\end{gather}
Variation of the action $S$ with respect to the Lagrange multipliers of course yields the set of constraints. Variation of $S$ with respect to the fields $g_{\m \n}$, ${\stackrel{\smash{\circ}}{\Gamma}} ^\lambda _{\m \n}$, $\omega_{\m \n}$ and $\phi$ entails the equations of motion\footnote{Concerning the signs, we recall that the signature chosen for the metric is `mostly minus'.}
\begin{subequations}\label{eq:EOM1-3}
\begin{gather}
\frac{1}{\kappa} R^{\m \n} + g^{\m \n} \nabla _\lambda s^\lambda = - {\stackrel{\smash{\circ}}{T}} ^{\m \n} ,\label{eq:EOM1}\\
\delta_\lambda ^\n s^\m + 2 m^{\m \n \rho} \omega_{\lambda \rho} = t^{\m \n}_{\lambda } ,\label{eq:EOM2}\\
 \left( \frac{1}{2 \kappa} R +{\cal L}_{M} \right) \theta^{\n \m}= 2 \stackrel{\circ}{\nabla}_\lambda \! m^{\lambda \m \n} ,\label{eq:EOM3}
\end{gather}
\end{subequations}
and
\begin{gather}
0 = \frac{\pa {\cal L}_{M} }{\pa \phi} - \frac{1}{w} \pa_\m \left( w \frac{\pa {\cal L}_{M} }{\pa (\pa_\m \phi)} \right)
= \frac{\pa {\cal L}_{M} }{\pa \phi} - \frac{1}{\sqrt{|g|}} \pa_\m \left( \sqrt{|g|} \frac{\pa {\cal L}_{M} }{\pa (\pa_\m \phi)} \right)\nonumber\\
\hphantom{0}{} = \frac{\pa {\cal L}_{M} }{\pa \phi} - \nabla_\m \left( \frac{\pa {\cal L}_{M} }{\pa (\nabla_\m \phi)} \right) .
\end{gather}
In equation~\eqref{eq:EOM1}, $R^{\m \n}$ denotes the Ricci tensor associated to the metric $\underline{g}$ and ${\stackrel{\smash{\circ}}{T}} ^{\m \n} $ denotes the `symplectic EMT' which is related to the metric (or Einstein--Hilbert) EMT $T_{\rm EH} ^{\m \n} \equiv \frac{-2}{\sqrt{|g|}} \frac{\delta S_{M}}{ \delta g_{\m \n}}$ (satisfying the covariant conservation law $\nabla_\m T^{\m \n}_{\rm EH} =0$) as follows:
\begin{gather}\label{eq:EHEMT}
T_{\rm EH} ^{\m \n} = {\stackrel{\smash{\circ}}{T}} ^{\m \n} - g^{\m \n} {\cal L}_{M} , \qquad \mbox{with} \qquad
- \frac{1}{2} {\stackrel{\smash{\circ}}{T}} ^{\m \n} \equiv \frac{\pa {\cal L}_{M} }{\pa g_{\m \n}}
- \frac{1}{w} \pa_\lambda \left( w \frac{\pa {\cal L}_{M} }{\pa (\pa_\lambda g_{\m \n})} \right) .
\end{gather}

The set of equations~\eqref{eq:EOM1-3} can be combined in the following manner~\cite{Dobrski:2015emm}. By applying $\delta^\lambda _\n \stackrel{\circ}{\nabla} _\m$ to equation~\eqref{eq:EOM2} and then substituting equation~\eqref{eq:EOM3}, we get an expression for $\stackrel{\circ}{\nabla}_\m s^\m = {\nabla}_\m s^\m $: substitution of the latter into equation~\eqref{eq:EOM1} leads to the result
\begin{gather}\label{eq:EinsteinEq}
 R^{\m \n} - \frac{1}{2} R g^{\m \n} + \Lambda g^{\m \n}= - \kappa T_{\rm EH} ^{\m \n}, \qquad \mbox{with} \qquad \Lambda \equiv \frac{\kappa}{n} \stackrel{\circ}{\nabla} _\m t_\n ^{\m \n} .
\end{gather}
Here, we recognize Einstein's tensor $G^{\m \n} \equiv R^{\m \n} - \frac{1}{2} R g^{\m \n} $. By applying the covariant derivati\-ve~$\nabla_\m$ to equation~\eqref{eq:EinsteinEq} and by taking into account that $\nabla_\m G^{\m \n} =0 = \nabla_\m T_{\rm EH} ^{\m \n}$ as well as $\nabla_\m g^{\m \n} =0$, one concludes that $\Lambda$ is necessarily a constant. Henceforth, relation~\eqref{eq:EinsteinEq} is \emph{Einstein's field equation}
(for the metric tensor) including a cosmological constant~$\Lambda$. More precisely~\cite{Dobrski:2015emm}, the solutions of the equations of motion~\eqref{eq:EOM1-3} must include a metric $\underline{g}$ which solves Einstein's field equation~\eqref{eq:EinsteinEq} and, conversely, for each solution of equation~\eqref{eq:EinsteinEq} there exists a weakly compatible metric/Fedosov structure and Lagrange multipliers satisfying the set of equations~\eqref{eq:EOM1-3}. The symplectic data are determined by the set of equations~\eqref{eq:SympConn1}, \eqref{eq:SympConn2}, \eqref{eq:WCM} which imply relation~\eqref{eq:GpropW}, i.e., the proportionality of $w$ and $ \sqrt{|g|}$. As noted in~\cite{Dobrski:2015emm}, the action $S_g + S_{L}$ characterizing pure gravity (with a~cosmological constant) might be of interest for the canonical quantization since it does not involve the square root of the determinant of the metric.

\paragraph{Scalar field coupled to a weakly compatible metric/Fedosov structure:}
Let us assume that the complex scalar field $\Phi$ is coupled minimally to the metric tensor $(g_{\m \n})$. The matter field action functional then reads
\begin{align}\label{eq:actionMGS}
S_{M} \equiv \int_{M} \frac{\underline{\omega}^m}{m!} {\cal L}_{M} , \quad \mbox{with} \qquad{\cal L}_{M} (\Phi, \Phi^*, \underline{g}) =g^{\m \n} \pa_\m \Phi^* \pa_\n \Phi - {\rm m}^2\Phi^* \Phi - \frac{\cg}{2} ( \Phi^* \Phi)^2 .
\end{align}
Under a general coordinate transformation (diffeomorphism) $x \mapsto x' (x)$, the fields $\Phi$ and ${\cal L}_{M} $ transform as scalar fields, e.g., $\Phi ' (x' ) = \Phi (x)$. The argumentation presented in equations~\eqref{eq:EOM1-3}--\eqref{eq:EinsteinEq} then applies. In particular, the tensor ${\stackrel{\smash{\circ}}{T}} ^{\m \n} $ (as given in~equation~\eqref{eq:EHEMT}) now only involves the derivative with respect to $g_{\m \n}$, i.e., $ - \frac{1}{2} {\stackrel{\smash{\circ}}{T}} ^{\m \n} \! \equiv {\pa {\cal L}_{M} }/{\pa g_{\m \n}}$, and we obtain the usual, covariantly conserved EMT for a complex scalar field:
\begin{gather} \label{eq:EHEMTscal}
T_{\rm EH} ^{\m \n} = \pa^\m \Phi^* \pa^\n \Phi + \pa^\n \Phi^* \pa^\m \Phi- g^{\m \n} {\cal L}_{M} .
\end{gather}

Let us presently assume that we rescale the scalar field $\Phi$ as we did in~equation~\eqref{eq:FreeModelActions}, i.e., $\Phi \equiv \sqrt{w} \phi$. While the weight factor $w = \sqrt{\det (\omega_{\m \n})}$ was invariant
under the Poincar\'e transformations~\eqref{eq:PoincTrans} in flat space (due to the fact that $|\det ({\Lambda ^\m}_\n) |=1$), it is no longer invariant in curved space under general coordinate transformations: it rather transforms like the density~$\sqrt{|g|}$, i.e.,
\begin{gather*}
w' = |\lambda | w , \qquad \mbox{with} \qquad \lambda \equiv \det \left( \frac{\pa x ^\m}{\pa x ^{\prime \n}} \right) .
\end{gather*}
Thus, the field $\phi = \Phi / \sqrt{w}$ transforms like a scalar density of weight $1/2$ (i.e., the weight attributed to matter fields in Section~\ref{sec:FTNC}):
\begin{gather*}
\phi ' = |\lambda |^{-1/2} \phi .
\end{gather*}
By rewriting the scalar field ${\cal L}_{M} $ given in equation~\eqref{eq:actionMGS} as well as the associated, covariantly conserved EMT~\eqref{eq:EHEMTscal} in terms of the densities $\phi$, $\phi^*$, we obtain the result
\begin{subequations}
\begin{gather}
 {\cal L}_{M} = w \left[ g^{\m \n} \cd_\m \phi^* \cd_\n \phi - \mbox{m}^2 \phi^* \phi - \frac{\cg}{2} w ( \phi^* \phi)^2 \right], \label{eq:LafEMT1} \\
 T_{\rm EH} ^{\m \n} = w \big[\cd^\m \phi^* \cd^\n \phi + \cd^\n \phi^* \cd^\m \phi \big]- g^{\m \n} {\cal L}_{M} , \label{eq:LafEMT2}
\end{gather}
\end{subequations}
where $\cd_\m \phi = \frac{1}{\sqrt{w}} \pa_\m ( \sqrt{w} \phi) = \pa_\m \phi + \frac{1}{2} (\pa_\m \ln w) \phi$ and $\cd_\m \phi^* = (\cd_\m \phi)^*$. In the Lagrangian~\eqref{eq:LafEMT1} we note the appearance of a global factor $w$ (which implies that the action $\int_{M} \frac{\underline{\omega}^m}{m!} {\cal L}_{M} = \int_{M} {\rm d}^nx w^2 [ \cdots ]$ involves a global factor $w^2$) as well as of an extra factor~$w$ in the quartic term.

If one considers the weakly compatible metric/Fedosov structure (i.e., $w \propto \sqrt{|g|}$), one cannot directly recover a flat space model for the coupling of the field~$\phi$ to the symplectic tensor since~$\sqrt{|g|}$ reduces to unity in the flat space limit ($g_{\m \n} \leadsto \eta_{\m \n}$) while $w$ is an $x$-dependent function in flat space\footnote{This problem is reminiscent of the one of taking the commutative limit ($\theta^{\m \n} \to 0$) of noncommutative models involving $w = (\det (\theta^{\m \n}))^{-1/2}$: this issue was circumvented in~Section~\ref{sec:FTNC} by a~two step procedure where the parameters $\theta^{\m \n}$ were first assumed to be constant so that $w$ could be factored out of the integral.}. Let us ignore for the moment being the relation $w \propto \sqrt{|g|}$ and consider the following procedure to obtain a Lagrangian model in flat space:
 \begin{gather*}
S_{M} \equiv \int_{M} \frac{\underline{\omega}^m}{m!} {\cal L}_{M}\propto\int_{M} {\rm d}^nx \sqrt{|g|} {\cal L}_{M} \quad \leadsto \quad
S_{M}\Big{|} _{\underline{g} = \underline{\eta}} \equiv \int_{\br ^n} {\rm d}^nx {\cal L}_{M} \Big{|} _{\underline{g} = \underline{\eta}} .
\end{gather*}
Then, the curved space Lagrangian~\eqref{eq:LafEMT1} yields the flat space action functional
 \begin{gather} \label{eq:AdHocAction}
 S_{M}\Big{|} _{\underline{g} = \underline{\eta}}= \int_{\br ^n} {\rm d}^nx w \left[\cd^\m \phi^* \cd_\m \phi - \mbox{m}^2 \phi^* \phi - \frac{\cg}{2} w ( \phi^* \phi)^2 \right] .
\end{gather}
Due to the extra factor $w$ in the self-interaction term, this flat space model is different from the one that we discussed in~equation~\eqref{eq:ActScalSymp} (see also the equivalent expression~\eqref{eq:InteractActions}) and, more generally, in~equation~\eqref{eq:ActScalar} for flat noncommutative space. Indeed, for the self-interaction term of a noncommutative model associated to the action~\eqref{eq:AdHocAction} one could consider
\begin{gather*}
 - \frac{\cg}{2} \int_{\br ^n} {\rm d}^nx w \left[ \sqrt{w} ( \phi^* \star \phi) \right] \star \left[ \sqrt{w} ( \phi^* \star \phi) \right]= - \frac{\cg}{2} \int_{\br ^n} {\rm d}^nx w \left[ w ( \phi^* \star \phi)^2 \right] .
\end{gather*}
As a matter of fact, the factor $w$ being invariant under Poincar\'e transformations in the flat space case, one could then consider some general functions of~$w$ as coefficients in $ {\cal L}_{M} $. These alternative flat space models lead to energy-momentum balance equations which differ from those we encountered, e.g., the EMT associated to the matter field functional~\eqref{eq:AdHocAction} is locally conserved by construction.

\subsection{Curved noncommutative space-time}\label{sec:CurvedNCST}

The \emph{Fedosov star product deformation} of the field theoretic model~\eqref{eq:actionMS1}--\eqref{eq:actionMS2} has been studied by M.~Dobrski~\cite{Dobrski:2015emm} for the case of pure gravity so as to obtain a theory of noncommutative gravity which is generally covariant and independent of the symplectic background, the latter being dynamical. The first modification brought about the noncommutativity (which appears at order $\theta ^2$) is already quite complex and has been determined by using a Mathematica package for tensor calculus. The incorporation of matter fields is beyond the scope of the present work, but in view of the previous discussion concerning the coupling of matter fields to gravity, the matter action functionals that we studied for $M= \br^n$ should appear naturally. Here, we only outline the formulation for pure gravity in terms of the notation used so far and we comment on matter fields.

One of the pioneering works on deformation quantization is the one of B.~Fedosov who constructed a star product for a generic symplectic manifold (see~\cite{Fedosov:1996, Waldmann:2007} for an introduction to these topics). As a matter of fact, any star product on such a manifold is equivalent to a Fedosov star product, i.e., the equivalence class of the latter star product comprises all other ones. The approach of Fedosov amounts to a geometric extension of the Groenewold--Moyal quantization, the latter applying only to a symplectic vector space. Quite generally, the \emph{deformation quantization} on a given space~$M$ is a deformation of the product of functions defined on this space in the direction of the Poisson bracket of these functions. Since the Poisson bracket involves derivatives, the formulation of star products on manifolds calls for the introduction of a linear connection, the latter defining a covariant derivative\footnote{A different possibility~\cite{Aschieri:2005zs, Reshetikhin, Schenkel:2010sc, Jambor:2004kc} for obtaining globally well defined expressions for the higher order derivatives $\pa{_\m} \pa_{\n} \cdots f$ of a function on a manifold is to replace the partial derivatives $\pa_\m$ by Lie derivatives with respect to some vector fields~$X_a$: this amounts to the definition of star products in terms of \emph{Drinfeld twists} or to consider a so-called twisted manifold. This construction and its comparison with the approach based on a linear connection will be addressed elsewhere.}. 
On a~symplectic manifold, a~natural choice for such a connection $\stackrel{\circ}{\nabla}$ is the one which is torsion-free and symplectic, see equations~\eqref{eq:SympConn1} and~\eqref{eq:SympConn2}. Indeed, the Fedosov scheme (and other constructions which it inspired like the globalization of the local expression for star products on Poisson manifolds) relies on the introduction of such a connection. The final results are usually presented in a somewhat abstract form, but an explicit (iteratively determined) expansion for the Fedosov star product has recently been elaborated, see~\cite{Dobrski:2015emm} and references therein. For smooth complex-valued (i.e., scalar) functions $f$,~$g$ on the Fedosov manifold $(M, \underline{\omega}, \stackrel{\circ}{\nabla} )$, the \emph{Fedosov star product} reads
\begin{gather}\label{eq:FedStarFcts}
f\star_{S} g = fg + \tfrac {\ri}{2} \theta^{\mu\nu} \stackrel{\circ}{\nabla}_\mu f \stackrel{\circ}{\nabla}_\nu g- \tfrac{1}{8} \theta^{\rho \s} \theta^{\m \n}
\stackrel{\circ}{\nabla}_{(\rho} \stackrel{\circ}{\nabla}_{\m )} \! f \stackrel{\circ}{\nabla}_{(\s } \stackrel{\circ}{\nabla}_{\n )} g
+ \mathcal O\big(\theta^3\big) ,
\end{gather}
where the brackets denote symmetrization. The higher order terms in $\theta$ involve explicitly the curvature tensor ${\stackrel{\smash{\circ}}{R}} ^\rho _{\ \sigma \m \n}$ associated to the connection $\stackrel{\circ}{\nabla}$, e.g., see~\cite{Dobrski:2015emm} for the third order term. For a flat connection $\stackrel{\circ}{\nabla}$, i.e., ${\stackrel{\smash{\circ}}{R}} ^\rho _{\ \sigma \m \n} =0$, one can choose vanishing coefficients ${\stackrel{\smash{\circ}}{\Gamma}} ^\lambda _{\ \m \n} $ and Darboux local coordinates on $M$ (i.e., constant functions $\theta^{\mu\nu} $): the expression~\eqref{eq:FedStarFcts} then reduces to the Groenewold--Moyal star product.

However the Fedosov construction is more general: for a given vector bundle ${\cal E}$ over the base manifold $M$ (i.e., the fiber ${\cal E}_x$ over any $x\in M$ is a vector space), the Fedosov construction defines the associative \emph{deformation of the $($noncommutative$)$ product of matrices}, i.e., the deformation of the endomorphism bundle $\operatorname{End} {\cal E}$ over $M$. (The fiber $\left( \operatorname{End} {\cal E} \right)_x$ over $x\in M$ is the set of endomorphisms of the vector space ${\cal E}_x$ and any such endomorphism is given by a square matrix once a basis has been chosen for the vector space ${\cal E}_x$ at each $x$.) The deformation of the matrix product requires the introduction of a connection $ \stackrel{{\cal E}}{\nabla}$ on ${\cal E}$ which has to be added to $ \stackrel{\circ}{\nabla}$: thus we have a \emph{total connection}
\begin{gather*}
\hat{\nabla} \equiv{} \stackrel{\circ}{\nabla} \otimes \id + \id \otimes \stackrel{{\cal E}}{\nabla} ,
\end{gather*}
acting on the vector bundle $TM \otimes {\cal E}$ (where $TM$ denotes the tangent bundle of $M$). For the connection coefficients and the curvature associated to the connection $\stackrel{{\cal E}}{\nabla}$, we again have
local expressions of the form
\begin{gather*}
\stackrel{{\cal E}}{\nabla} _\m \equiv \pa_\m+ \underline{\stackrel{{\cal E}}{\Gamma}_\m} , \qquad \mbox{with} \qquad
\underline{\stackrel{{\cal E}}{\Gamma}_\m}\equiv \big( \stackrel{\smash{\cal E}}{\Gamma} ^{\lambda} _{\m \n} \big) ,
\qquad \underline{\stackrel{{\cal E}}{R}_{\m \n}} \equiv \pa_\m \underline{\stackrel{{\cal E}}{\Gamma}_\n}
- \pa_\n \underline{\stackrel{{\cal E}}{\Gamma}_\m} + \big[ \underline{\stackrel{{\cal E}}{\Gamma}_\m} , \underline{\stackrel{{\cal E}}{\Gamma}_\n} \big] ,
\end{gather*}
and analogously for the total connection $\hat{\nabla}$. The \emph{Fedosov star product of $x$-dependent matrices $($endomorphisms$)$} $\underline{F} \equiv (F^\m_{\ \n})$ and $\underline{G} \equiv (F^\m_{\ \n})$ now reads~\cite{Dobrski:2010pu, Dobrski:2015emm}
\begin{gather}
\underline{F} \star \underline{G} = \underline{F} \underline{G} + \tfrac {\ri}{2} \theta^{\mu\nu} \hat{\nabla}_\mu \underline{F} \hat{\nabla}_\nu \underline{G}- \tfrac{1}{8} \theta^{\rho \s} \theta^{\m \n} \Big(\hat{\nabla}_{(\rho} \hat{\nabla}_{\m )} \underline{F} \hat{\nabla}_{(\s } \hat{\nabla}_{\n )} \underline{G}\nonumber\\
\hphantom{\underline{F} \star \underline{G} =}{}
+ \big\{ \hat{\nabla}_{\sigma } \underline{F} , \underline{\stackrel{{\cal E}}{R} _{\rho \m}} \big\} \hat{\nabla}_{\n } \underline{G} + \hat{\nabla}_{\sigma } \underline{F} \big\{ \underline{\stackrel{{\cal E}}{R} _{\rho \m}} , \hat{\nabla}_{\n } \underline{G} \big\} \Big) + \mathcal O\big(\theta^3\big) ,\label{eq:FedStarMatr}
\end{gather}
where $\{, \cdot , \cdot \}$ denotes the anticommutator of matrices. For a flat connection $\stackrel{{\cal E}}{\nabla}$, i.e., for $\underline{\stackrel{{\cal E}}{R}_{\m \n}} =0$, and the choice $\underline{\stackrel{{\cal E}}{\Gamma}_\m} =0$, the star product~\eqref{eq:FedStarMatr} reduces to a star product of matrices for which the multiplication of entries is given by the star product of functions~\eqref{eq:FedStarFcts}. We note that the Fedosov star product \eqref{eq:FedStarMatr} is not closed~\cite{Dobrski:2010pu}.

For the formulation of field theories and more precisely of action functionals, it is again necessary to introduce an appropriate cyclic trace functional $\operatorname{tr}_\star$, i.e., $\operatorname{tr}_\star (\underline{F} \star \underline{G} ) = \operatorname{tr}_\star (\underline{G} \star \underline{F} )$ for compactly supported endomorphisms $\underline{F}$, $\underline{G}$. Such a functional has also been introduced by B.~Fedosov and an explicit expression for it has been worked out by this author in~\cite{Fedosov2002} (see also~\cite{Dobrski:2010pu} where a Mathematica package for tensor calculus is applied):
\begin{gather}\label{eq:FedTrace}
\operatorname{tr}_\star \underline{F} \equiv \int_{M} \frac{\underline{\omega}^m}{m!} \operatorname{Tr} \big[\underline{F} - \tfrac{\ri}{2} \theta^{\mu\nu} \underline{\stackrel{{\cal E}}{R}_{\m \n}} \underline{F}
+ \mathcal O\big(\theta^2\big) \big] .
\end{gather}
For the formulation of \emph{pure gravity}, the Fedosov manifold $ (M , \underline{\omega}, \stackrel{{\circ}}{\Gamma}) $ endowed with a given metric structure, one considers the vector bundle ${\cal E} \equiv TM$ and the Levi-Civita connection on this bundle, i.e., $ \stackrel{{\cal E}}{\nabla} = \nabla$ (see equations~\eqref{eq:LeviCivita1} and~\eqref{eq:LeviCivita2}). Then, the natural choice for an endomorphism $\underline{F} \in \operatorname{End} TM$ in an action functional of the type~\eqref{eq:FedTrace} is given by the Ricci tensor with the first index raised, i.e., $\underline{F} \equiv \underline{{\cal R}}$ with $ \underline{{\cal R}} \equiv (R^\m_{\ \n}) \equiv (R^{\rho \m}_{\ \ \n \rho}) $. Indeed, with this choice one has the real, diffeomorphism invariant action functional
\begin{gather}\label{eq:ActNCGrav}
S_\txt{ncg} \equiv\frac{1}{2\kappa} \operatorname{tr}_\star \underline{{\cal R}} = S_g +S_\txt{nc} .
\end{gather}
Here, the first term is the functional $S_g$ considered for pure gravity in commutative space-time (see equation~\eqref{eq:actionMS1}) and~$S_\txt{nc} $ represents the noncommutative corrections. The latter are of order~$\theta^2$ and higher order in~$\theta$ since the term in~\eqref{eq:FedTrace} which is linear in $\theta$ vanishes for $\underline{F} = \underline{{\cal R}}$ due to the symmetry properties of the curvature tensor:
$\operatorname{Tr} \big( \underline{\stackrel{{\cal E}}{R}_{\m \n}} \underline{{\cal R}}\big) = R^{\rho}_{\ \sigma \m \n} R^{\lambda \sigma}_{\ \ \rho \lambda} =0$. This appears to be a general feature of noncommutative gravity~\cite{Dobrski:2015emm}.

The calculation~\eqref{eq:actionMS1}--\eqref{eq:EinsteinEq} can now be generalized by starting from the complete action functional $S_\txt{ncg} + S_{L}$ with $S_\txt{ncg}$ given by~\eqref{eq:ActNCGrav} and $S_{L}$ given by~\eqref{eq:actionMS2}. This leads to a set of equations of the form~\eqref{eq:EOM1-3} with $\stackrel{\smash{\circ}}{T} ^{\m \n} =0 = {\cal L}_{M}$ and with a noncommutative correction term in each equation. Explicit expressions for the latter terms up to order $\theta^2$ have been given in~\cite{Dobrski:2015emm}. The elimination of the Lagrange multipliers is presently more complex than in the commutative case due to the appearance of integrability conditions. In view of the complexity which already underlies pure noncommutative gravity, the incorporation of matter fields (which also requires the introduction of tetrad fields in the case of Dirac spinors) is beyond the scope of the present work.

\subsection{Particular examples of curved noncommutative space-time}\label{sec:ExCurvedNCST}
Rather than studying the dynamics of space-time, we can also choose a given curved space-time (e.g., four-dimensional space-time endowed with the Schwarzschild metric) and study deformed field theories on such a background.
A simple, but non trivial example for a Riemannian manifold is given by an orientable surface. For such a space-time of dimension $n=2$, the Poisson tensor~$(\theta ^{\m \n})$ only involves a single independent component $\theta ^{12} \big(x^1, x^2\big) \equiv \vartheta \big(x^1, x^2\big) $ which implies that $w \equiv \sqrt{\det (\omega_{\m \n})} = 1/\sqrt{ \det (\theta^{\m \n})} = 1/|\vartheta |$. By virtue of~equation~\eqref{eq:GpropW}, the compatibility of the metric and Poisson structures then implies that $1/|\vartheta |$ is proportional to $\sqrt{g}$ where $g$ denotes the determinant of the metric tensor~$(g_{\m \n})$. For simplicity, we consider $\vartheta$ to be positive and equal to $1/\sqrt{g}$ \cite{Beggs:2017yzq}, i.e.,
\begin{gather*}
 (\omega _{\m \n} ) =
 \left[
 \begin{matrix}
 0 & -w
 \\
 w & 0
 \end{matrix}
 \right] , \qquad \mbox{with} \qquad w = \sqrt{g} .
\end{gather*}
Thus the symplectic volume (area) form~\eqref{eq:VolElem}, i.e., ${\rm d}V_{\omega} = \underline{\omega} = \frac{1}{2} \omega_{\m \n} {\rm d}x^\m \wedge {\rm d}x^\n = - w {\rm d}x^1 \wedge {\rm d}x^2$ and the Riemannian volume (area) form ${\rm d}V_g = \sqrt{g} {\rm d}x^1 \wedge {\rm d}x^2$ coincide with each other up to the sign. The Weyl star product formula~\eqref{eq:KontStarProduct} applied to the coordinates $x^1$, $x^2$ then implies $\big[ x^1 \stackrel{\star}{,} x^2 \big] = \ri h \vartheta \big(x^1, x^2 \big)$ where we spelled out the formal deformation parameter~$h$.

Let us now suppose for concreteness that the orientable surface under consideration has constant curvature, i.e., it is (up to a homeomorphism) a $2$-sphere for the case of positive curvature or the hyperbolic plane for the case of negative curvature. We will elaborate briefly on the example of the \emph{unit $2$-sphere}~\cite{Beggs:2017yzq}. For the latter the upper hemisphere can be parametrized by Cartesian coordinates $\big(x^1, x^2, x^3\big) \equiv (x, y,z )$ with $x^2+y^2 <1$ and $z = z(x,y) \equiv \sqrt{1-x^2 -y^2}$ (and similarly for the lower hemisphere). In terms of these coordinates, the standard line element ${\rm d}s ^2 = {\rm d}\theta ^2 + \sin^2 \theta {\rm d}\varphi^2$ of the $2$-sphere is given by
\begin{gather*}
{\rm d}s^2 \equiv g_{\m \n} {\rm d}x^\m {\rm d}x^\n = \frac{1}{z^2} \big[ \big(1 -y^2\big) {\rm d}x^2+ 2xy {\rm d}x {\rm d}y + \big(1 -x^2\big) {\rm d}y^2\big] .
\end{gather*}
This entails that $\sqrt{g} = 1/z$, hence we have coordinate dependent components for the Poisson tensor which are given by $\theta^{12} = z(x,y) = - \theta^{21}$. The Weyl star product formula~\eqref{eq:KontStarProduct} or the closed star product~\eqref{eq:StarProduct} then imply
\begin{gather*}
\big[ x^i \stackrel{\star}{,} x^j \big] = \ri h \varepsilon^{ijk} x^k + {\cal O} \big(h^2\big) , \qquad \mbox{with} \quad i, j \in \{ 1,2,3 \} .
\end{gather*}
At the first order in $h$, these star-commutation relations are the ones which characterize the \emph{fuzzy sphere}. The Levi-Civita connection (associated to the standard metric) and the symplectic two-form considered to be the (opposite of the) Riemannian volume form give the structure of a~Fedosov manifold to the $2$-sphere~\cite{Gelfand}.

The case of the hyperbolic plane modeled by the Poincar\'e upper halfplane \mbox{$\big\{ (x,y) \!\in\! \br^2 \,|\, y>0 \big\}$} endowed with the metric ${\rm d}s^2 = y^{-2} \big({\rm d}x^2 + {\rm d}y^2\big)$ can be studied along the same lines and leads to the results $w = 1/y^2$ and $[ x \stackrel{\star}{,} y ] = \ri h y^2$. These considerations again fit into the general framework discussed before, yet the dynamical study of matter fields on these spaces obviously requires some further work.

\section{Concluding remarks}\label{sec:Conclusion}

\paragraph{About the quantum theory:} For $\br_{\theta}^3$, i.e., $\br^3$ with $\textbf{su}(2)$-noncommutativity (in which case the commutator algebra $\big[\hat X^i , \hat X^j \big] = \ri \varepsilon^{ijk} \hat X^k$ which may be realized by the Pauli matrices), the one-loop quantization of self-interacting scalar field theory has recently been investigated~\cite{Juric:2016cfp}, in particular for the scalar field model described by the classical action~\eqref{eq:ActScalar} (with $w\equiv 1$ for $\br_{\theta}^3$). The main results may be summarized as follows~\cite{Juric:2016cfp} (see also~\cite{Poulain:2018mcm} for some further recent work). The $2$-point function does not involve infrared singularities in the external momenta (even in the massless case) which indicates the absence of the infamous UV/IR mixing problem for these models. This result appears to have its origin in the Lie algebraic nature of the underlying noncommutativity. Moreover, the $2$-point function is finite in the ultraviolet regime where the deformation parameter corresponds to an ultraviolet cut-off $\Lambda \propto 1/\theta$.

\paragraph{Conclusion:} The mathematical framework for field theories on a space-time defined by generic noncommutativity parameters is more complex than the one for constant parameters. Nevertheless the classical theory can be formulated to a large extent along similar lines. In this context, some interesting mathematical structures appear which may be worthwhile to explore further. The simple approach to the conservation laws that we considered
here is also of interest in other contexts and its application to gauge field theories, supersymmetry, conformal models, etc.\ will be discussed elsewhere. The presented analogy of free field models with the damped harmonic oscillator is intriguing and may also be useful for the investigation of some aspects of the quantum theory. For the latter one also has to tackle the subtleties of time ordering in the {\nc} setting, see, e.g.,~\cite{Bahns:2002} and references therein for a discussion on this point in Moyal space. A scheme for describing the dynamics of the fields $\theta^{\m \n}(x)$ was outlined in curved space-time, but an elaboration and better understanding of this point definitely requires further work. The case where the matrix $(\theta^{\m \n}(x))$ does not have maximal rank for all~$x$ (e.g., $\theta^{12}$ vani\-shing on a line in the two-dimensional case) also requires extra work involving a regularization of integrals, e.g., see~\cite{Fosco:2004jb}: this instance appears to be of interest for the study of boundary effects occurring in condensed matter systems.

\appendix

\section{Simple construction of star products}\label{sec:ConsStarProd}

\paragraph{General star product:} In this appendix we indicate how explicit expressions for a star product can be obtained in a simple manner. By way of motivation, we start from the case of constant noncommutativity parameters $\theta^{\m \n}$ and the corresponding commutation relations in quantum mechanics:
\begin{gather}\label{eq:canNCQMalgebra}
\big[ \hat X ^\m , \hat X ^\n \big] = \ri \theta^{\m \n} \id , \qquad
\big[ \hat P _\m , \hat P _\n \big] = 0 , \qquad
\big[ \hat X ^\m , \hat P _\n \big] = - \ri \delta ^\m _\n \id .
\end{gather}
 Here, the operators $\hat X ^\m$ and $\hat P_\n$ are supposed to act on a complex separable Hilbert space. A~representation of the algebra~\eqref{eq:canNCQMalgebra} in terms of the standard operators of position~$X^\m$ (which acts on wave functions as multiplication by the real variable $x^\m$) and of momentum $P_\n \equiv \ri \pa_\n$ satisfying canonical commutation relations is well known~\cite{ Chaichian:2000si, Delduc:2007av, Mezincescu:2000zq}:
 \begin{gather}\label{eq:RepresCanNCQM}
\hat X ^\m = X^\m + \tfrac{1}{2} \theta^{\m \n} P_\n = X^\m + \tfrac{\ri}{2} \theta^{\m \n} \pa_\n , \qquad \hat P _\n = P_\n = \ri \pa_\n .
\end{gather}
An operator function $\hat f$ of the operators $\hat X^\m$ (defined by using the Weyl ordering prescription) then acts on a smooth function $g$ by means of the Groenewold--Moyal star product~\cite{Bigatti:1999iz, Mezincescu:2000zq}:
\begin{gather*}
\hat f\big(\hat X\big) g = \hat f \big( X^\m + \tfrac{\ri}{2} \theta^{\m \n} \pa_\n \big) g
= fg + \tfrac{\ri}{2} \theta^{\m \n} \pa_\m f \pa_\n g +\tfrac{1}{2} \big( \tfrac{\ri}{2} \big)^2 \theta^{\m \n} \theta^{\rho \sigma} \pa_\m \pa_\rho f \pa_\n \pa_\sigma g + \cdots .
\end{gather*}

For general noncommutativity parameters $\theta^{\m \n} (x) = - \theta^{\n \m} (x)$, a representation of the commutation relations~\eqref{eq:one_operators}, i.e., of $\big[ \hat X ^\m , \hat X ^\n \big] = \ri \hat{\theta} ^{\m \n} \big(\hat X \big)$, can be found by generalizing the expression~\eqref{eq:RepresCanNCQM} of~$\hat X^\m$ or, more precisely, by expanding the operators $\hat X^\m$ and $\hat{\theta} ^{\m \n} \big(\hat X \big) $ as polydifferential operators while considering the Weyl ordering prescription. This procedure has been worked out by Kupriyanov and Vassilevich~\cite{Kupriyanov:2008dn} (see also~\cite{Kupriyanov:2012rf}) so as to determine an explicit expression for the star product up to fourth order in $\theta$. The result to first order in $\theta$ already follows by considering the expression~\eqref{eq:RepresCanNCQM} for~$\hat X^\m$ in terms of $x$-dependent parameters $\theta^{\m \n}$, i.e., $\hat X ^\m = X^\m + \frac{\ri}{2} \theta^{\m \n} \pa_\n$: one finds that
\begin{gather*}
\big[ \hat X ^\m , \hat X ^\n \big] = \ri \hat{\theta} ^{\m \n} \big(\hat X \big) ,
\end{gather*}
with
\begin{gather}
 \hat{\theta} ^{\m \n} \big(\hat X \big) = \hat{\theta} ^{\m \n} \big( X^\sigma + \tfrac{\ri}{2} \theta^{\sigma \rho} \pa_\rho + {\cal O} \big(\theta ^2 \big) \big)
 = \theta^{\m \n} + \tfrac{\ri}{2} \theta^{\sigma \rho } \pa _\sigma \theta^{\m \n} \pa_\rho + {\cal O} \big(\theta ^3 \big) ,\label{eq:RepNCalg}
\end{gather}
where the operator $ \hat{\theta} ^{\m \n} ( X )$ acts as usual as multiplication by the function $ \theta^{\m \n} (x)$. Moreover, one readily checks that the Jacobi identity for commutators (i.e., the relation
$0 = \big[ \hat X ^\lambda , \big[ \hat X ^\m , \hat X ^\n \big] \big] + \mbox{cyclic permutations of the indices} \ \lambda,\, \m,\, \n$) is satisfied to order $\theta$ if the antisymmetric tensor $\theta^{\m \n}$ satisfies the Poisson--Jacobi identity characterizing a Poisson tensor. Finally, the linear term of the star product also follows straightforwardly:
 \begin{gather*}
f\star g = \hat f \big(\hat X\big) g = \hat f \big( X^\m + \tfrac{\ri}{2} \theta^{\m \n} \pa_\n + {\cal O} \big(\theta ^2\big) \big) g = fg + \tfrac{\ri}{2} \theta^{\m \n} \pa_\m f \pa_\n g + {\cal O} \big(\theta ^2 \big) .
\end{gather*}
For the terms of higher order in $\theta$, one has to expand $\hat X ^\m$ and $\hat{\theta}^{\m \n}$ to higher order than first, which yields~\cite{Kupriyanov:2012rf, Kupriyanov:2008dn}
\begin{gather}\label{eq:Xpolydiff}
\hat X ^\m = X^\m + \tfrac{\ri}{2} \theta^{\m \n} \pa_\n + \tfrac{1}{24} \big[ \theta^{\rho \sigma} \pa_\sigma \theta^{\m \n} + \theta^{\nu \sigma} \pa_\sigma \theta^{\m \rho} \big] \pa_\rho \pa_\n + {\cal O} \big(\theta^3 \big) ,
\end{gather}
as well as an expression for $ \hat{\theta} ^{\m \n} \big(\hat X\big) $ (as a differential polynomial in the Poisson tensor $\theta^{\m \n}$) which extends the first order result~\eqref{eq:RepNCalg}. For the definition of a Weyl-ordered function $\hat f\big( \hat X \big) $, one applies the general formula~\cite{Berezin, Kupriyanov:2015uxa, Szabo:2001}
 \begin{gather*}
 \hat f\big(\hat X\big) = \frac{1}{(2\pi )^n} \int_{\br^n} {\rm d}^nk \tilde{f} (k) \re^{-{\rm i} k_\m \hat X^\m} ,
 \end{gather*}
where $\tilde f(k) = \int_{\br^n} {\rm d}^nx {f} (x) \re^{{\rm i} k_\m x^\m}$ denotes the Fourier transform of $f$. The latter relation defines the Weyl symbol of the operator $\hat f\big(\hat X\big) $. The expression for the star product which results to order $\theta^2$ from this procedure is given by equation~\eqref{eq:KontStarProduct} and is commented upon in that context.

\paragraph{Closed star product:} For physical applications, we are interested in a closed star product~\cite{Kupriyanov:2012rf}, i.e., such that relation~\eqref{eq:Closure} holds for some integration measure $w d^nx$.
As we noted in equation~\eqref{eq:ClosRelOrder1}, the closure relation for the star product yields (at the first order in $\theta$) the divergenceless condition $\pa_\m (w \theta ^{\m \n})=0$. By taking into account the latter as well as the Jacobi identity for the Poisson tensor and by performing some integrations by parts, the closure relation leads, to the second order in $\theta$, to the result
\begin{gather}\label{eq:ClosRelOrder2}
0= \int_{\br ^n} {\rm d}^nx w ( f \star g - fg ) = \tfrac{1}{24} \int_{\br ^n} {\rm d}^nx B^{\rho \nu} \pa_\rho f \pa_\n g .
 \end{gather}
Here, the matrix $(B^{\rho \nu} )$ whose elements are given by $B^{\rho \nu} = \pa_\m ( w \theta^{\rho \sigma} \pa_\sigma \theta^{\m \n} ) $ is symmetric by virtue of the divergenceless condition and the Jacobi identity for the Poisson tensor $(\theta^{\m \n})$. The result~\eqref{eq:ClosRelOrder2} means that the Weyl star product does not satisfy the closure relation. However, this result suggests to make a judicious equivalence (or so-called gauge) transformation~\cite{Kontsevich:1997vb} of the star product $\star$ so as to obtain a~star product~$\star '$ which is closed to order~$\theta ^2$ (see~\cite{Kupriyanov:2012rf} for the treatment of~\eqref{eq:ClosRelOrder2} and~\cite{Kupriyanov:2015uxa} for the general procedure to all orders in~$\theta$): by considering a linear differential operator of the form $D = \id + d^{\rho \nu}
 \pa_\rho \pa_\n + {\cal O}\big(\theta^3\big)$, one readily finds that the \emph{gauge transformed star product} $\star '$ induced by $D$ has the following form (to order~$\theta^2$):
\begin{gather*}
f \star ' g \equiv D^{-1} ( Df \star Dg ) = f \star g - 2 d^{\rho \nu} \pa_\rho f \pa_\n g .
\end{gather*}
Thus the particular choice $2 d^{\rho \nu} = \frac{1}{24 w} B^{\rho \n}$ allows to eliminate the nonvanishing term on the right hand side of equation~\eqref{eq:ClosRelOrder2}, i.e., to obtain a gauge equivalent star product $\star '$ which satisfies the closure relation $\int_{\br ^n} {\rm d}^nx w f \star ' g = \int_{\br ^n} {\rm d}^nx w f g$ to order~$\theta ^2$. Its explicit expression (up to order~$\theta^2$) is spelled out in equation~\eqref{eq:StarProduct} where we suppressed the prime on the star product. The basic operator $\hat X^\m$ now becomes $\hat X ^{\prime \m}$ and its $\theta$-expansion follows from $\hat X ^{\prime \m} g \equiv x^\m \star ' g$: in comparison to the operator $\hat X ^\m$ given by expression~\eqref{eq:Xpolydiff} it thus involves, at order $\theta^2$, a supplementary contribution. The latter ensures~\cite{Kupriyanov:2012rf, Kupriyanov:2013jka} that the operator $ \hat X ^{\prime \m}$ is Hermitian with respect to the inner product~\eqref{eq:defSP}.

\section{Some comments on Sections~\ref{sec:ncspace} and~\ref{sec:FTNC}}\label{sec:MathComment}

In this appendix, we gather some mathematical remarks concerning Sections~\ref{sec:ncspace} and~\ref{sec:FTNC}, respectively.

\paragraph{About the star product approach:} Concerning the mathematical framework, we note that the transformation of the volume form ${\rm d}^nx \leadsto w {\rm d}^nx$ with $w = \sqrt{ \det (\omega_{\m \n}) } $ represents a~resca\-ling. Although the general relationship between the definition of noncommutative geometry by A.~Connes in terms of spectral triples and the construction of noncommutative spaces by deformations of commutative algebras (which we follow here) is not completely clear~\cite{Connes:2006ms} (see however~\cite{Gayral:2003dm} for some results), the rescaling ${\rm d}^nx \leadsto w {\rm d}^nx$ appears to be related to the so-called twisted spectral triples introduced in~\cite{ConnesMoscovici} and further studied in~\cite{Moscovici}. Indeed, an example for the latter is given by the gauge transformed spectral triple induced by a rescaling of the Riemannian metric.

\paragraph{The case of two space-time dimensions:} In~\cite{Kupriyanov:2008dn}, it was pointed out that simplifications should occur for the star products in two space-time dimensions, i.e., for $n=2$. In this respect we only note~\cite{Ovsienko} that one can pass over from $\br^2 \setminus \{ (0,0) \}$ parametrized by the variables \mbox{$\big(x^0 \equiv p$}, $x^1 \equiv q\big)$ to the circle (i.e., real projective space $\br \mathbb{P}^1$) parametrized by the single va\-riab\-le \mbox{$\xi \equiv q/p$}: in this case, one identifies a \emph{tensor density} $\xi \mapsto \phi (\xi)$ of degree $\lambda \in \mathbb{R}$ on $\br \mathbb{P}^1$ with the \emph{homogeneous function} $F_{\phi}\colon \br^2 \setminus \{ (0,0) \} \to \mathbb{C}$ of degree $-2 \lambda$ given by
\begin{gather*}
(p,q) \longmapsto F_{\phi} (p,q) \equiv p^{-2\lambda} \phi \left( \frac{q}{p} \right) .
\end{gather*}
The Groenewold--Moyal star product $F_{\phi} \star F_{\psi}$ can then be written as a star product of the densities $\phi$, $\psi$ (of degree~$\lambda$ and~$\m$ respectively) and represents a power series in the real variable $\vartheta \equiv \theta^{01}$:
\begin{gather}\label{eq:TensDens}
\phi \star \psi = \phi \psi + \sum_{k=1}^\infty \left( \frac{\ri \vartheta }{2} \right)^k J^{\lambda, \mu}_{k} (\phi ,\psi ) .
\end{gather}
Here, the quantities $J^{\lambda, \mu}_{k}$ are the \emph{Gordan transvectants} which are well-known bilinear differential operators that are invariant under projective (i.e., M\"obius or fractional-linear) transformations and which appear for instance in classical $W$-algebras and conformal models, e.g., see~\cite{Gieres:2000rw} and references therein. In particular~\cite{Ovsienko}, the Gordan transvectant $J^{\lambda, \mu}_{1} (\phi ,\psi ) \propto \lambda \phi \psi ' - \mu \phi ' \psi$ represents the so-called \emph{Schouten bracket} of $\phi$ and $\psi$. The star product~\eqref{eq:TensDens} is invariant under the group $\textbf{PGL} (2, \br)$ of projective transformations which is homomorphic to the group $\textbf{Sp}(2, \br)$ of symplectic transformations. The star product in two dimensions can also be generalized to Riemann surfaces, e.g., see~\cite{Biswas}.

\paragraph{About the modified Leibniz rule:} Consider the associative, commutative algebra ${\cal A} \equiv C^{\infty} (\br ^n)$ of smooth functions on $\br ^n$ and write the elements of the deformation matrix as \mbox{$\theta^{\m \n} = h \Theta ^{\m \n}$} where the real parameter~$h$ represents a formal deformation parameter. Furthermore, let~${\cal A} [[ h ]]$ denote the algebra of formal power series $\sum\limits_{n=0}^\infty h^n a_n$ with coefficients $a_n$ in ${\cal A}$, equipped with the star product. The fact that the linear operator $\cd_\m\colon {\cal A} [[ h ]] \to {\cal A} [[ h ]]$ introduced in equation~\eqref{eq:defCD} does not satisfy the Leibniz rule (i.e., $\cd_\m$ does not represent a $\star$-derivation) is equivalent to the statement that the operator $T \equiv \re^{h a^\m \cd_\m}$ (with $a^\m \in \br$) does not represent a~$\star$-automorphism, i.e., $T(f\star g) \neq (Tf) \star (Tg)$~-- see~\cite[Proposition~6.2.7]{Waldmann:2007}.

Linear operators on associative algebras which do not satisfy the Leibniz rule appear in various contexts in physics and in mathematics, e.g., in the Batalin--Vilkovisky anti-bracket formulation of gauge field theories~\cite{YKS:1996, ClaudeRoger, Witten:1990wb} or in general relativity, see~\cite{Coll:2003ym} and references therein. The deviation from the Leibniz rule of a linear operator (acting on an associative algebra like ${\cal A} [[ h ]]$ equipped with the star product) is known in mathematics as the \emph{Hochschild differential} $\delta$ of this operator~\cite{Waldmann:2007}: for the operator $\cd_\m$, the bilinear map $\delta \cd_\m \colon {\cal A} [[ h ]] \otimes {\cal A} [[ h ]] \to {\cal A} [[ h ]]$ is given by
\begin{gather*}
-(\delta \cd_\m ) (f,g) = \cd_\m (f \star g) - (\cd_\m f) \star g - f \star (\cd_\m g) .
\end{gather*}
According to relation~\eqref{eq:LeibnizRule}, we thereby have
\[
(\delta \cd_\m ) (f,g)
= w^{-1} \pa_\rho b^{\rho}_\m (f,g)
= \cd_{\rho} B^{\rho}_\m (f,g)
\qquad \mbox{with} \qquad
B^{\rho}_\m (f,g) \equiv w^{-1} b^{\rho}_\m (f,g) .
\]
Thus, the Hochschild differential of $\cd_\m$ looks like a coboundary term. In the physics literature, the deviation from the Leibniz rule has also been qualified as the Leibniz bracket~\cite{Coll:2003ym}. More precisely, for the algebra ${\cal A} [[ h ]]$ equipped with the star product and the grading introduced after~equation~\eqref{ex:ExpDerivStar}, the \emph{Leibniz bracket of the linear operator $\cd_{\m}$ with respect to the star product} is defined (in terms of the notation of~\cite{Coll:2003ym}) by
\begin{gather*}
\{ f, g \} _{\cd _\m} \equiv {\cal L} _{\cd _\m} \langle \star \rangle (f,g)\equiv \cd _\m f \star g + f \star \cd _\m g - \cd_\m (f\star g) .
\end{gather*}
We will not elaborate on these mathematical aspects here, but their application should be worth exploring in greater detail. We only mention that the fundamental property of the Leibniz bracket is given by
\begin{gather*}
{\cal L} _{[ \cd _\m, \cd_\n ]} \langle \star \rangle= {\cal L} _{\cd _\n} \langle {\cal L} _{\cd _\m} \langle \star \rangle \rangle- {\cal L} _{\cd _\m} \langle {\cal L} _{\cd _\n} \langle \star \rangle \rangle .
\end{gather*}

\section{Damped harmonic oscillator}\label{sec:dampedosci}

The models of free Lagrangian field theories on noncommutative space discussed in this paper and the corresponding energy-momentum conservation laws admit close analogies with the Lagrangian formulation of a damped harmonic oscillator in non-relativistic mechanics and with a~corresponding conserved quantity. Therefore, we describe the latter system in this appendix by stressing the analogies using an appropriate choice of notation.

For a mechanical system with one degree of freedom, i.e., a single second order differential equation $\ddot{q}= f (q, \dot{q} , t)$ for the particle's position $t \mapsto q(t)$, it has already been shown by Jacobi and Darboux that a Lagrangian function always exists. A simple example for a~dissipative system in one dimension is given by the \emph{damped harmonic oscillator}, i.e., a dynamical system governed by a~differential equation with constant coefficients $m>0$, $k>0$, $\gamma \geq 0$:
\begin{gather*}
m \ddot{q} + k q + \gamma m \dot q =0 ,
\end{gather*}
or, with $\omega \equiv \sqrt{k/m}$,
\begin{gather}\label{dho}
\ddot{q} + \omega^2 q + \gamma \dot q =0 .
\end{gather}
To simplify the notation, we will consider a unit mass in the following. As we just stated, a single differential equation of second order like (\ref{dho}) can always be obtained as an Euler--Lagrange equation, eventually after multiplying it by an \emph{integrating multiplier}, i.e., a non-vanishing function $t \mapsto \wt (t)$ in the present one-dimensional case. Indeed~\cite{Bateman:1931,Caldirola:1941,Kanai:1948}, the Lagrangian
\begin{gather}\label{eq:LagDOH}
L (q , \dot q , t) = \wt \left[ \frac{1}{2} \dot q ^2 - \frac{\omega^2}{2} q ^2 \right] , \qquad \mbox{with} \qquad \wt (t) \equiv \txt{e}^{\gamma t},
\end{gather}
which describes a harmonic oscillator with time dependent mass and stiffness (or frequency), yields the Euler--Lagrange equation
\begin{gather*}
0 = \frac{\partial L }{ \partial q} - \frac{{\rm d} }{{\rm d}t} \left( \frac{\partial L }{ \pa \dot q}\right) = - \wt \big[ \ddot{q} + \omega^2 q + \gamma \dot q \big] ,
\end{gather*}
i.e., the equation of motion (\ref{dho}). Since $\gamma = \pa_t \ln \wt$, the last equation can also be written as
\begin{gather*}
0 = \frac{1}{\wt} \pa_t (\wt \dot q ) + \omega^2 q ,
\end{gather*}
and thus has the same structure as the equations of motion~\eqref{eq:FreeFEphi} of our field theoretical models for free scalar and Dirac fields in noncommutative space.

For the discussion of the conservation law below, we spell out the solution of the equation of motion~\eqref{dho} which satisfies given initial conditions $q(0) = x_0$ and $\dot q (0) = v_0$: with $\Omega^2 \equiv \omega^2 - \big( \frac{\gamma}{2} \big)^2$, we have
\begin{gather}\label{eq:SolEOM}
q(t) = \re^{-\frac{\gamma}{2} t} \left[ x_0 \cos \Omega t + \frac{1}{\Omega} \left( v_0 + \frac{\gamma}{2} x_0 \right)\sin \Omega t \right] .
\end{gather}

We note that the canonical momentum associated to $q$ is given by $p \equiv \partial L /\pa \dot q = \wt \dot q$, hence the canonical Hamiltonian reads
\begin{gather}\label{eq:HamDHO}
H (q , p , t) \equiv p \dot q - L= \wt ^{-1} \frac{p^2}{2} + \wt \frac{\omega^2}{2} q ^2 .
\end{gather}
This function may also be expressed in terms of $q$, $\dot q$ and $t$,
\begin{gather*}
H= \wt H_0 , \qquad \mbox{with} \qquad H_0 \equiv \frac{1}{2} \dot q ^2 + \frac{\omega^2}{2} q ^2 ,
\end{gather*}
where $H_0$ represents the total energy of the undamped oscillator. However, due to the dissipation, the Hamiltonian $H$ is not a conserved quantity: we have an \emph{energy balance} equation which can be determined straightforwardly by using the equation of motion for~$q$:
\begin{gather*}
\frac{{\rm d}H_0}{{\rm d}t} = - \gamma \dot q ^2= - (\pa_t \ln \wt ) \dot q ^2 .
\end{gather*}

Nevertheless, a \emph{conserved charge} can be constructed by different methods and in particular as follows by performing some redefinitions~\cite{ClassMech}. By virtue of the time-dependent rescaling
\begin{gather*} 
q \leadsto Q = \wt ^{1/2} q
\end{gather*}
(which is analogous to the $x$-dependent rescaling of matter fields encountered for our field theoretical models in equation~\eqref{eq:FreeModelActions}), the equation of motion for $q$ takes the form of the one for an undamped oscillator:
\begin{gather*}
\ddot Q + \Omega^2 Q = 0 , \qquad \mbox{with} \qquad \Omega^2 \equiv \omega^2 - \left( \frac{\gamma}{2} \right)^2 .
\end{gather*}
For the latter dynamical system parametrized by $Q$, the total energy $E$ is obviously conserved and is given by
\begin{gather}\label{eq:ConsCharge}
2E \equiv \dot Q^2 + \Omega^2 Q^2 = \txt{e}^{\gamma t} \big[ \dot q^2 + \omega^2 q^2 + \gamma q \dot q \big] .
\end{gather}
Here the last expression is the rewriting of the conserved charge in terms of the original variable~$q$. It represents an \emph{explicitly time-dependent conserved quantity} for the dissipative system under consideration. Its physical interpretation can be elucidated by considering its value at the time $t=0$:
\begin{gather*}
2E (t) = 2 E(0) = v_0^2 + \omega^2 x_0 ^2 + \gamma x_0 v_0 .
\end{gather*}
Thus, the conserved quantity is simply a particular combination of the initial conditions, the first two terms representing the energy of the undamped oscillator. In the course of the motion, the exponentially increasing
 factor in the charge~\eqref{eq:ConsCharge} is compensated by the exponential decrease of the solution~\eqref{eq:SolEOM} of the equation of motion.

Within the Hamiltonian formulation, the conserved quantity $E$ (expressed in terms of the phase space variables~$q$,~$p$) generates local transformations $\delta q$, $\delta p$ of the phase space variables by means of the Poisson brackets and, conversely, the latter transformations give rise to the conserved charge $E$ by virtue of Noether's first theorem. More precisely, with the standard Poisson bracket of functions $F$, $G$ on phase space,
\begin{gather*}
\{ F, G \} \equiv \frac{\pa F}{\pa q} \frac{\pa G}{\pa p} - \frac{\pa F}{\pa p} \frac{\pa G}{\pa q} ,
\end{gather*}
we find
\begin{gather}
\delta q \equiv \{ q , E \} = \dot{q} + \frac{\gamma}{2} q = \big[ \pa_t + \big(\pa_t \ln \wt ^{1/2} \big) \big] q , \nonumber\\
\delta p \equiv \{ p , E \} = \dot{p} - \frac{\gamma}{2} p = \big[ \pa_t - \big(\pa_t \ln \wt ^{1/2} \big) \big] p.\label{eq:SymTransfDHO}
\end{gather}
Hence the variation of the Lagrangian~\eqref{eq:LagDOH} under these transformations reads
\begin{gather}\label{eq:DivSym}
\delta L = \frac{{\rm d}L}{{\rm d}t} .
\end{gather}
Thus, we have a divergence symmetry of the action, namely $\delta L = \frac{{\rm d}f}{{\rm d}t} $ with $f = L$. According to Noether's first theorem the conserved charge associated to such a divergence symmetry reads
\begin{gather*}
{\cal E} = \frac{\pa L}{\pa \dot q} \delta q - L ,
\end{gather*}
where the last factor describes the divergence symmetry~\eqref{eq:DivSym}. Substitution of the expression for $L$ yields ${\cal E} =E$.

We conclude with two comments concerning the conservation laws in a dissipative dynamical system like the damped harmonic oscillator. First, we note that non-autonomous Hamiltonian systems like the one given by the explicitly time-dependent Hamiltonian~\eqref{eq:HamDHO} can be described in a \emph{symplectic extended phase space}, see~\cite{Struckmeier:2005} for a general study. In that framework, a time-dependent Hamiltonian can be mapped by a generalized canonical transformation into a time-independent Hamiltonian. Indeed, our conserved charge~\eqref{eq:ConsCharge} coincides with the invariant~(46) of~\cite{Struckmeier:2005} (upon considering $n=1$, $\omega = \txt{const}$, $F(t) = \gamma t$, and $\xi =1$ in~\cite{Struckmeier:2005}).

Second, we remark that for a linear, explicitly time-dependent dynamical system like the damped harmonic oscillator, one can perform a so-called \emph{Arnold transformation} \cite[Section~1.6.A]{Arnold} which maps the equation of motion of the system into the one of a free particle by virtue of a mapping $(q,t) \mapsto (\xi, \tau )$ of the underlying non-relativistic space-time into itself, see~\cite{Cariglia:2016oft} and references therein. Indeed, by decomposing the general solution~\eqref{eq:SolEOM} of the equation of motion subject to the initial conditions $q(0) = x_0$ and $\dot q (0) = v_0$ as
\begin{gather*}
q(t) = v_0 u_1 (t) + x_0 u_2 (t) \qquad \mbox{with} \qquad
\begin{cases}
\displaystyle u_1(t) = \re^{-\frac{\gamma}{2} t} \frac{1}{\Omega} \sin \Omega t,\vspace{1mm}\\
\displaystyle u_2(t) = \re^{-\frac{\gamma}{2} t} \left( \cos \Omega t + \frac{\gamma}{2\Omega} \sin \Omega t \right) ,
\end{cases}
\end{gather*}
we obtain the solution $\tau \mapsto \xi (\tau)$ of the free particle equation of motion,
\begin{gather*}
\xi (\tau) = v_0 \tau + x_0
 \qquad \mbox{with} \qquad \begin{cases}
\displaystyle \xi \equiv \frac{q}{u_2} = \frac{\re^{\frac{\gamma}{2} t} q}{\cos \Omega t + \frac{\gamma}{2\Omega} \sin \Omega t}, \vspace{1mm}\\
\displaystyle \tau \equiv \frac{u_1}{u_2} = \frac{\sin \Omega t}{\Omega \cos \Omega t + \frac{\gamma}{2} \sin \Omega t} .
\end{cases}
\end{gather*}
An extension of the two-dimensional space-time, $(q, t) \leadsto (q,t,s)$, to a three-dimensional space (which is referred to as \emph{Bargmann space} or \emph{Eisenhart lift}) and the related extension $ (q,t,s) \mapsto (\xi, \tau, \sigma)$ of the Arnold map then allows to show~\cite{Cariglia:2016oft} that the damped harmonic oscillator has the same symmetries as the free particle (and thereby also admits corresponding conserved quantities
 whose expression can be derived by means of the extended Arnold map).

\subsection*{Acknowledgments}

The authors are indebted to H.~Grosse, C.~Roger, V.~Kupriyanov, M.~Dobrski and H.~Steinacker for reading a preliminary version of our manuscript and for their instructive remarks on some of the mathematical aspects.
We are also grateful to the anonymous referees for their important remarks. We sincerely thank D.~Grumiller, K.~Landsteiner and A.~Rebhan for organizing a~memorial meeting in the memory of Manfred Schweda at the Vienna University of Technology and for inviting us to present our joint work with Manfred on that occasion.

\pdfbookmark[1]{References}{ref}
\LastPageEnding


\begin{thebibliography}{99}
\footnotesize\itemsep=-0.4pt

\bibitem{Arefeva:2000uu}
Aref'eva I.Ya., Belov D.M., Koshelev A.S., Rytchkov O.A., Renormalizability and
 {UV}/{IR} mixing in noncommutative theories with scalar fields, \href{https://doi.org/10.1016/S0370-2693(00)00831-5}{\textit{Phys.
 Lett.~B}} \textbf{487} (2000), 357--365.

\bibitem{Arnold}
Arnold V.I., Geometrical methods in the theory of ordinary differential
 equations, 2nd ed., \href{https://doi.org/10.1007/978-1-4612-1037-5}{\textit{Grundlehren der Mathematischen Wissenschaften}}, Vol.~250,
Springer-Verlag, New York, 1988.

\bibitem{Aschieri:2005yw}
Aschieri P., Blohmann C., Dimitrijevi\'{c} M., Meyer F., Schupp P., Wess J., A
 gravity theory on noncommutative spaces, \href{https://doi.org/10.1088/0264-9381/22/17/011}{\textit{Classical Quantum Gravity}}
 \textbf{22} (2005), 3511--3532, \href{https://arxiv.org/abs/hep-th/0504183}{hep-th/0504183}.

\bibitem{Aschieri:2009zz}
Aschieri P., Dimitrijevi\'{c} M., Kulish P., Lizzi F., Wess J., Noncommutative
 spacetimes. Symmetries in noncommutative geometry and field theory,
 \href{https://doi.org/10.1007/978-3-540-89793-4}{\textit{Lecture Notes in Phys.}}, Vol.~774, Springer-Verlag, Berlin, 2009.

\bibitem{Aschieri:2005zs}
Aschieri P., Dimitrijevi\'{c} M., Meyer F., Wess J., Noncommutative geometry
 and gravity, \href{https://doi.org/10.1088/0264-9381/23/6/005}{\textit{Classical Quantum Gravity}} \textbf{23} (2006),
 1883--1911, \href{https://arxiv.org/abs/hep-th/0510059}{hep-th/0510059}.

\bibitem{Bagchi:2009wb}
Bagchi B., Fring A., Minimal length in quantum mechanics and non-{H}ermitian
 {H}amiltonian systems, \href{https://doi.org/10.1016/j.physleta.2009.09.054}{\textit{Phys. Lett.~A}} \textbf{373} (2009),
 4307--4310, \href{https://arxiv.org/abs/0907.5354}{arXiv:0907.5354}.

\bibitem{Bahns:2002}
Bahns D., Doplicher S., Fredenhagen K., Piacitelli G., On the unitarity problem
 in space/time noncommutative theories, \href{https://doi.org/10.1016/S0370-2693(02)01563-0}{\textit{Phys. Lett.~B}} \textbf{533}
 (2002), 178--181, \href{https://arxiv.org/abs/hep-th/0201222}{hep-th/0201222}.

\bibitem{Balasin:2015hna}
Balasin H., Blaschke D.N., Gieres F., Schweda M., On the energy-momentum tensor
 in {M}oyal space, \href{https://doi.org/10.1140/epjc/s10052-015-3492-8}{\textit{Eur. Phys.~J.~C Part. Fields}} \textbf{75} (2015),
 284, 11~pages, \href{https://arxiv.org/abs/1502.03765}{arXiv:1502.03765}.

\bibitem{Banerjee:2009gr}
Banerjee R., Chakraborty B., Ghosh S., Mukherjee P., Samanta S., Topics in
 noncommutative geometry inspired physics, \href{https://doi.org/10.1007/s10701-009-9349-y}{\textit{Found. Phys.}} \textbf{39}
 (2009), 1297--1345, \href{https://arxiv.org/abs/0909.1000}{arXiv:0909.1000}.

\bibitem{Bateman:1931}
Bateman H., On dissipative systems and related variational principles,
 \href{https://doi.org/10.1103/PhysRev.38.815}{\textit{Phys. Rev.}} \textbf{38} (1931), 815--819.

\bibitem{Bayen:1977ha}
Bayen F., Flato M., Fronsdal C., Lichnerowicz A., Sternheimer D., Deformation
 theory and quantization. {I}.~{D}eformations of symplectic structures,
 \href{https://doi.org/10.1016/0003-4916(78)90224-5}{\textit{Ann. Physics}} \textbf{111} (1978), 61--110.

\bibitem{Beggs:2017yzq}
Beggs E.J., Majid S., Poisson--{R}iemannian geometry, \href{https://doi.org/10.1016/j.geomphys.2016.12.012}{\textit{J.~Geom. Phys.}}
 \textbf{114} (2017), 450--491, \href{https://arxiv.org/abs/1403.4231}{arXiv:1403.4231}.

\bibitem{Behr:2003qc}
Behr W., Sykora A., Construction of gauge theories on curved noncommutative
 spacetime, \href{https://doi.org/10.1016/j.nuclphysb.2004.07.024}{\textit{Nuclear Phys.~B}} \textbf{698} (2004), 473--502,
 \href{https://arxiv.org/abs/hep-th/0309145}{hep-th/0309145}.

\bibitem{Behr:2003hg}
Behr W., Sykora A., N{C} {W}ilson lines and the inverse {S}eiberg--{W}itten map
 for non-degenerate star products, \href{https://doi.org/10.1140/epjc/s2004-01778-4}{\textit{Eur. Phys.~J.~C Part. Fields}}
 \textbf{35} (2004), 145--148, \href{https://arxiv.org/abs/hep-th/0312138}{hep-th/0312138}.

\bibitem{Geloun:2007zza}
Ben~Geloun J., Hounkonnou M.N., Noncommutative {N}oether theorem, in X{XVI}
 {W}orkshop on {G}eometrical {M}ethods in {P}hysics, \href{https://doi.org/10.1063/1.2820980}{\textit{AIP Conf. Proc.}},
 Vol.~956, Amer. Inst. Phys., Melville, NY, 2007, 55--60.

\bibitem{Berezin}
Berezin F.A., Shubin M.A., The {S}chr\"{o}dinger equation, \href{https://doi.org/10.1007/978-94-011-3154-4}{\textit{Mathematics
 and its Applications (Soviet Series)}}, Vol.~66, Kluwer Academic Publ. Group, Dordrecht, 1991.

\bibitem{Bieliavsky}
Bieliavsky P., Cahen M., Gutt S., Rawnsley J., Schwachh\"{o}fer L., Symplectic
 connections, \href{https://doi.org/10.1142/S021988780600117X}{\textit{Int.~J. Geom. Methods Mod. Phys.}} \textbf{3} (2006),
 375--420, \href{https://arxiv.org/abs/math.SG/0511194}{math.SG/0511194}.

\bibitem{Bigatti:1999iz}
Bigatti D., Susskind L., Magnetic fields, branes, and noncommutative geometry,
 \href{https://doi.org/10.1103/PhysRevD.62.066004}{\textit{Phys. Rev.~D}} \textbf{62} (2000), 066004, 6~pages,
 \href{https://arxiv.org/abs/hep-th/9908056}{hep-th/9908056}.

\bibitem{Biswas}
Biswas I., Differential operators on a {R}iemann surface with projective
 structure, \href{https://doi.org/10.1016/j.geomphys.2003.09.002}{\textit{J.~Geom. Phys.}} \textbf{50} (2004), 393--414.

\bibitem{Blaschke:2016ohs}
Blaschke D.N., Gieres F., Reboud M., Schweda M., The energy-momentum tensor(s)
 in classical gauge theories, \href{https://doi.org/10.1016/j.nuclphysb.2016.07.001}{\textit{Nuclear Phys.~B}} \textbf{912} (2016),
 192--223.

\bibitem{Blaschke:2010kw}
Blaschke D.N., Kronberger E., Sedmik R.I.P., Wohlgenannt M., Gauge theories on
 deformed spaces, \href{https://doi.org/10.3842/SIGMA.2010.062}{\textit{SIGMA}} \textbf{6} (2010), 062, 70~pages,
 \href{https://arxiv.org/abs/1004.2127}{arXiv:1004.2127}.

\bibitem{Blumenhagen:2018kwq}
Blumenhagen R., Brunner I., Kupriyanov V., L\"{u}st D., Bootstrapping
 non-commutative gauge theories from {$\rm L_\infty$} algebras,
 \href{https://doi.org/10.1007/JHEP05(2018)097}{\textit{J.~High Energy Phys.}} \textbf{2018} (2018), no.~5, 097, 46~pages,
 \href{https://arxiv.org/abs/1803.00732}{arXiv:1803.00732}.

\bibitem{Caldirola:1941}
Caldirola P., Forze non conservative nella meccanica quantistica, \href{https://doi.org/10.1007/BF02960144}{\textit{Nuovo
 Cim.}} \textbf{18} (1941), 393--400.

\bibitem{Calmet:2005qm}
Calmet X., Kobakhidze A., Noncommutative general relativity, \href{https://doi.org/10.1103/PhysRevD.72.045010}{\textit{Phys.
 Rev.~D}} \textbf{72} (2005), 045010, 5~pages, \href{https://arxiv.org/abs/hep-th/0506157}{hep-th/0506157}.

\bibitem{Calmet:2003jv}
Calmet X., Wohlgenannt M., Effective field theories on noncommutative
 space-time, \href{https://doi.org/10.1103/PhysRevD.68.025016}{\textit{Phys. Rev.~D}} \textbf{68} (2003), 025016, 11~pages,
 \href{https://arxiv.org/abs/hep-ph/0305027}{hep-ph/0305027}.

\bibitem{Cariglia:2016oft}
Cariglia M., Duval C., Gibbons G.W., Horv\'{a}thy P.A., Eisenhart lifts and
 symmetries of time-dependent systems, \href{https://doi.org/10.1016/j.aop.2016.07.033}{\textit{Ann. Physics}} \textbf{373}
 (2016), 631--654, \href{https://arxiv.org/abs/1605.01932}{arXiv:1605.01932}.

\bibitem{Cattaneo:1999fm}
Cattaneo A.S., Felder G., A path integral approach to the {K}ontsevich
 quantization formula, \href{https://doi.org/10.1007/s002200000229}{\textit{Comm. Math. Phys.}} \textbf{212} (2000),
 591--611, \href{https://arxiv.org/abs/math.QA/9902090}{math.QA/9902090}.

\bibitem{Chaichian:2010yt}
Chaichian M., Oksanen M., Tureanu A., Zet G., Covariant star product on
 symplectic and {P}oisson space-time manifolds, \href{https://doi.org/10.1142/S0217751X10049785}{\textit{Internat.~J. Modern
 Phys.~A}} \textbf{25} (2010), 3765--3796, \href{https://arxiv.org/abs/1001.0503}{arXiv:1001.0503}.

\bibitem{Chaichian:2010yw}
Chaichian M., Oksanen M., Tureanu A., Zet G., Noncommutative gauge theory using
 a covariant star product defined between {L}ie-valued differential forms,
 \href{https://doi.org/10.1103/PhysRevD.81.085026}{\textit{Phys. Rev.~D}} \textbf{81} (2010), 085026, 11~pages,
 \href{https://arxiv.org/abs/1001.0508}{arXiv:1001.0508}.

\bibitem{Chaichian:2000si}
Chaichian M., Sheikh-Jabbari M.M., Tureanu A., Hydrogen atom spectrum and the
 {L}amb shift in noncommutative {QED}, \href{https://doi.org/10.1103/PhysRevLett.86.2716}{\textit{Phys. Rev. Lett.}} \textbf{86}
 (2001), 2716--2719, \href{https://arxiv.org/abs/hep-th/0010175}{hep-th/0010175}.

\bibitem{Chamseddine:2003cw}
Chamseddine A.H., Noncommutative gravity, \href{https://doi.org/10.1007/s00023-003-0968-0}{\textit{Ann. Henri Poincar\'{e}}}
 \textbf{4} (2003), S881--S887, \href{https://arxiv.org/abs/hep-th/0301112}{hep-th/0301112}.

\bibitem{Chamseddine:2010ps}
Chamseddine A.H., Connes A., Space-time from the spectral point of view, in
 Recent Developments in Theo\-retical and Experimental General Relativity,
 Astrophysics and Relativistic Field Theories, \mbox{Vols.~1--3}, Proceedings of the
 12th Marcel Grossmann Meeting (Paris, July 12--18, 2009), Editors T.~Damour,
 R.~Jantzen, R.~Ruffini, \href{https://doi.org/10.1142/9789814374552_0001}{World Scientific, Singapore}, 2012, 3--23,
 \href{https://arxiv.org/abs/1008.0985}{arXiv:1008.0985}.

\bibitem{Ciric:2018ygk}
\'Ciri\'c~Dimitrijevi\'c M., Go\u{c}anin D., Konjik N., Radovanovi\'c V.,
 Noncommutative electrodynamics from {${\rm SO}(2,3)_\star$} model of
 noncommutative gravity, \href{https://doi.org/10.1140/epjc/s10052-018-6015-6}{\textit{Eur. Phys.~J.~C Part. Fields}} \textbf{78}
 (2018), 548, 13~pages, \href{https://arxiv.org/abs/1804.00608}{arXiv:1804.00608}.

\bibitem{Coll:2003ym}
Coll B., Ferrando J.J., On the {L}eibniz bracket, the {S}chouten bracket and
 the {L}aplacian, \href{https://doi.org/10.1063/1.1738188}{\textit{J.~Math. Phys.}} \textbf{45} (2004), 2405--2410,
 \href{https://arxiv.org/abs/gr-qc/0306102}{gr-qc/0306102}.

\bibitem{ConnesFlatoSternheimer}
Connes A., Flato M., Sternheimer D., Closed star products and cyclic
 cohomology, \href{https://doi.org/10.1007/BF00429997}{\textit{Lett. Math. Phys.}} \textbf{24} (1992), 1--12.

\bibitem{Connes:2006ms}
Connes A., Marcolli M., A walk in the noncommutative garden, in An Invitation
 to Noncommutative Geo\-met\-ry, Editors M.~Khalkhali, M.~Marcolli, \href{https://doi.org/10.1142/9789812814333_0001}{World Sci.
 Publ.}, Hackensack, NJ, 2008, 1--128, \href{https://arxiv.org/abs/math.QA/0601054}{math.QA/0601054}.

\bibitem{ConnesMoscovici}
Connes A., Moscovici H., Type {III} and spectral triples,
 \href{https://arxiv.org/abs/math.OA/0609703}{math.OA/0609703}.

\bibitem{Cornalba:2001sm}
Cornalba L., Schiappa R., Nonassociative star product deformations for
 {D}-brane world-volumes in curved backgrounds, \href{https://doi.org/10.1007/s002201000569}{\textit{Comm. Math. Phys.}}
 \textbf{225} (2002), 33--66, \href{https://arxiv.org/abs/hep-th/0101219}{hep-th/0101219}.

\bibitem{Das:2003kw}
Das A., Frenkel J., Kontsevich product and gauge invariance, \href{https://doi.org/10.1103/PhysRevD.69.065017}{\textit{Phys.
 Rev.~D}} \textbf{69} (2004), 065017, 6~pages, \href{https://arxiv.org/abs/hep-th/0311243}{hep-th/0311243}.

\bibitem{Delduc:2007av}
Delduc F., Duret Q., Gieres F., Lefrancois M., Magnetic fields in
 noncommutative quantum mechanics, \href{https://doi.org/10.1088/1742-6596/103/1/012020}{\textit{J.~Phys. Conf. Ser.}} \textbf{103}
 (2008), 012020, 26~pages, \href{https://arxiv.org/abs/0710.2239}{arXiv:0710.2239}.

\bibitem{Deriglazov:2017jub}
Deriglazov A.A., Ram\'{i}rez W.G., Recent progress on the description of
 relativistic spin: vector model of spinning particle and rotating body with
 gravimagnetic moment in general relativity, \href{https://doi.org/10.1155/2017/7397159}{\textit{Adv. Math. Phys.}}
 (2017), 7397159, 49~pages, \href{https://arxiv.org/abs/1710.07135}{arXiv:1710.07135}.

\bibitem{Dito:2002dr}
Dito G., Sternheimer D., Deformation quantization: genesis, developments and
 metamorphoses, in Deformation Quantization ({S}trasbourg, 2001), \textit{IRMA
 Lect. Math. Theor. Phys.}, Vol.~1, de Gruyter, Berlin, 2002, 9--54,
 \href{https://arxiv.org/abs/math.QA/0201168}{math.QA/0201168}.

\bibitem{Dobrski:2010pu}
Dobrski M., On some models of geometric noncommutative general relativity,
 \href{https://doi.org/10.1103/PhysRevD.84.065005}{\textit{Phys. Rev.~D}} \textbf{84} (2011), 065005, 11~pages,
 \href{https://arxiv.org/abs/1011.0165}{arXiv:1011.0165}.

\bibitem{Dobrski:2015emm}
Dobrski M., Background independent noncommutative gravity from {F}edosov
 quantization of endomorphism bundle, \href{https://doi.org/10.1088/1361-6382/aa5f82}{\textit{Classical Quantum Gravity}}
 \textbf{34} (2017), 075004, 17~pages, \href{https://arxiv.org/abs/1512.04504}{arXiv:1512.04504}.

\bibitem{Esposito:2015}
Esposito C., Formality theory. From Poisson structures to deformation
 quantization, \href{https://doi.org/10.1007/978-3-319-09290-4}{\textit{SpringerBriefs in Mathematical Physics}}, Vol.~2,
 Springer, Cham, 2015.

\bibitem{Fedosov:1996}
Fedosov B., Deformation quantization and index theory, \textit{Mathematical
 Topics}, Vol.~9, Akademie Verlag, Berlin, 1996.

\bibitem{Fedosov2002}
Fedosov B., On the trace density in deformation quantization, in Deformation
 Quantization ({S}trasbourg, 2001), \href{https://doi.org/10.1007/978-3-0348-9092-2_14}{\textit{IRMA Lect. Math. Theor. Phys.}},
 Vol.~1, de Gruyter, Berlin, 2002, 67--83.

\bibitem{Felder:2000nc}
Felder G., Shoikhet B., Deformation quantization with traces, \href{https://doi.org/10.1023/A:1026577414320}{\textit{Lett.
 Math. Phys.}} \textbf{53} (2000), 75--86, \href{https://arxiv.org/abs/math.QA/0002057}{math.QA/0002057}.

\bibitem{Fosco:2004jb}
Fosco C.D., Torroba G., Planar field theories with space-dependent
 noncommutativity, \href{https://doi.org/10.1088/0305-4470/38/16/016}{\textit{J.~Phys.~A: Math. Gen.}} \textbf{38} (2005),
 3695--3707, \href{https://arxiv.org/abs/hep-th/0405090}{hep-th/0405090}.

\bibitem{Fring:2010pw}
Fring A., Gouba L., Scholtz F.G., Strings from position-dependent
 noncommutativity, \href{https://doi.org/10.1088/1751-8113/43/34/345401}{\textit{J.~Phys.~A: Math. Theor.}} \textbf{43} (2010),
 345401, 10~pages, \href{https://arxiv.org/abs/1003.3025}{arXiv:1003.3025}.

\bibitem{Fritz:2016svs}
Fritz C., Majid S., Noncommutative spherically symmetric spacetimes at
 semiclassical order, \href{https://doi.org/10.1088/1361-6382/aa72a5}{\textit{Classical Quantum Gravity}} \textbf{34} (2017),
 135013, 50~pages, \href{https://arxiv.org/abs/1611.04971}{arXiv:1611.04971}.

\bibitem{Gayral:2003dm}
Gayral V., Gracia-Bond\'{i}a J.M., Iochum B., Sch\"{u}cker T., V\'{a}rilly
 J.C., Moyal planes are spectral triples, \href{https://doi.org/10.1007/s00220-004-1057-z}{\textit{Comm. Math. Phys.}}
 \textbf{246} (2004), 569--623, \href{https://arxiv.org/abs/hep-th/0307241}{hep-th/0307241}.

\bibitem{Gayral:2005ih}
Gayral V., Gracia-Bond\'{i}a J.M., Ruiz~Ruiz F., Position-dependent
 noncommutative products: classical construction and field theory,
 \href{https://doi.org/10.1016/j.nuclphysb.2005.08.016}{\textit{Nuclear Phys.~B}} \textbf{727} (2005), 513--536,
 \href{https://arxiv.org/abs/hep-th/0504022}{hep-th/0504022}.

\bibitem{Gelfand}
Gelfand I., Retakh V., Shubin M., Fedosov manifolds, \href{https://doi.org/10.1006/aima.1998.1727}{\textit{Adv. Math.}}
 \textbf{136} (1998), 104--140, \href{https://arxiv.org/abs/dg-ga/9707024}{dg-ga/9707024}.

\bibitem{Schweda:2000}
Gerhold A., Grimstrup J., Grosse H., Popp L., Schweda M., Wulkenhaar R., The
 energy-momentum tensor on noncommutative spaces: some pedagogical comments,
 \href{https://arxiv.org/abs/hep-th/0012112}{hep-th/0012112}.

\bibitem{Gieres:2000rw}
Gieres F., Conformal covariance in {$2D$} conformal and integrable models, in $W$-algebras and in their supersymmetric extensions,
in Proceedings of 3rd International Workshop on Supersymmetries and Quantum Symmetries (Dubna, July 1999),
2000, \href{https://arxiv.org/abs/nlin.SI/0002036}{nlin.SI/0002036}.

\bibitem{Gomes:2009tk}
Gomes M., Kupriyanov V.G., Position-dependent noncommutativity in quantum
 mechanics, \href{https://doi.org/10.1103/PhysRevD.79.125011}{\textit{Phys. Rev.~D}} \textbf{79} (2009), 125011, 6~pages.

\bibitem{Gomes:2009rz}
Gomes M., Kupriyanov V.G., da~Silva A.J., Dynamical noncommutativity,
 \href{https://doi.org/10.1088/1751-8113/43/28/285301}{\textit{J.~Phys.~A: Math. Theor.}} \textbf{43} (2010), 285301, 9~pages,
 \href{https://arxiv.org/abs/0908.2963}{arXiv:0908.2963}.

\bibitem{GraciaBondia:2010sm}
Gracia-Bond{\'{i}}a J.M., Notes on `quantum gravity' and non-commutative
 geometry, in New Paths Towards Quantum Gravity, \href{https://doi.org/10.1007/978-3-642-11897-5_1}{\textit{Lect. Notes Phys.}},
 Vol.~807, Editors B.~Booss-Bavnbek, G.~Esposito, M.~Lesch, Springer-Verlag,
 Berlin, 2002, 3--58, \href{https://arxiv.org/abs/1005.1174}{arXiv:1005.1174}.

\bibitem{Grensing:2013}
Grensing G., Structural aspects of quantum field theory and noncommutative
 geometry, \href{https://doi.org/10.1142/8771}{World Scientific Publ.}, Hackensack, NJ, 2013.

\bibitem{Groenewold:1946}
Groenewold H.J., On the principles of elementary quantum mechanics,
 \href{https://doi.org/10.1007/978-94-017-6065-2_1}{\textit{Physica}} \textbf{12} (1946), 1--56.

\bibitem{Grosse:2007jr}
Grosse H., Wohlgenannt M., Renormalization and induced gauge action on a
 noncommutative space, \href{https://doi.org/10.1143/PTPS.171.161}{\textit{Progr. Theoret. Phys. Suppl.}} \textbf{171}
 (2007), 161--177, \href{https://arxiv.org/abs/0706.2167}{arXiv:0706.2167}.

\bibitem{Hawkins:2002rf}
Hawkins E., Noncommutative rigidity, \href{https://doi.org/10.1007/s00220-004-1036-4}{\textit{Comm. Math. Phys.}} \textbf{246}
 (2004), 211--235, \href{https://arxiv.org/abs/math.QA/0211203}{math.QA/0211203}.

\bibitem{Hawkins:2007}
Hawkins E., The structure of noncommutative deformations,
 \href{https://doi.org/10.4310/jdg/1193074900}{\textit{J.~Differential Geom.}} \textbf{77} (2007), 385--424,
 \href{https://arxiv.org/abs/math.QA/0504232}{math.QA/0504232}.

\bibitem{Herbst:2001ai}
Herbst M., Kling A., Kreuzer M., Star products from open strings in curved
 backgrounds, \href{https://doi.org/10.1088/1126-6708/2001/09/014}{\textit{J.~High Energy Phys.}} \textbf{2001} (2001), no.~9, 014,
 21~pages, \href{https://arxiv.org/abs/hep-th/0106159}{hep-th/0106159}.

\bibitem{Hohm:2017pnh}
Hohm O., Zwiebach B., {$L_\infty$} algebras and field theory, \href{https://doi.org/10.1002/prop.201700014}{\textit{Fortschr.
 Phys.}} \textbf{65} (2017), 1700014, 33~pages, \href{https://arxiv.org/abs/1701.08824}{arXiv:1701.08824}.

\bibitem{Horvathy:2010wv}
Horv\'{a}thy P.A., Martina L., Stichel P.C., Exotic {G}alilean symmetry and
 non-commutative mechanics, \href{https://doi.org/10.3842/SIGMA.2010.060}{\textit{SIGMA}} \textbf{6} (2010), 060, 26~pages,
 \href{https://arxiv.org/abs/1002.4772}{arXiv:1002.4772}.

\bibitem{Juric:2016cfp}
Juri\'{c} T., Poulain T., Wallet J.-C., Closed star product on noncommutative
 {${\mathbb R}^3$} and scalar field dynamics, \href{https://doi.org/10.1007/JHEP05(2016)146}{\textit{J.~High Energy Phys.}}
 \textbf{2016} (2016), no.~5, 146, 22~pages, \href{https://arxiv.org/abs/1603.09122}{arXiv:1603.09122}.

\bibitem{Kanai:1948}
Kanai E., On the quantization of the dissipative systems, \href{https://doi.org/10.1143/ptp/3.4.440}{\textit{Rep. Progr.
 Phys.}} \textbf{3} (1948), 440--442.

\bibitem{Kontsevich:1997vb}
Kontsevich M., Deformation quantization of {P}oisson manifolds, \href{https://doi.org/10.1023/B:MATH.0000027508.00421.bf}{\textit{Lett.
 Math. Phys.}} \textbf{66} (2003), 157--216, \mbox{\href{https://arxiv.org/abs/q-alg/9709040}{q-alg/9709040}}.

\bibitem{YKS:1996}
Kosmann-Schwarzbach Y., From {P}oisson algebras to {G}erstenhaber algebras,
 \href{https://doi.org/10.5802/aif.1547}{\textit{Ann. Inst. Fourier (Grenoble)}} \textbf{46} (1996), 1243--1274.

\bibitem{Kupriyanov:2012nb}
Kupriyanov V.G., A hydrogen atom on curved noncommutative space,
 \href{https://doi.org/10.1088/1751-8113/46/24/245303}{\textit{J.~Phys.~A: Math. Theor.}} \textbf{46} (2013), 245303, 7~pages,
 \href{https://arxiv.org/abs/1209.6105}{arXiv:1209.6105}.

\bibitem{Kupriyanov:2012rf}
Kupriyanov V.G., Quantum mechanics with coordinate dependent noncommutativity,
 \href{https://doi.org/10.1063/1.4830032}{\textit{J.~Math. Phys.}} \textbf{54} (2013), 112105, 25~pages,
 \href{https://arxiv.org/abs/1204.4823}{arXiv:1204.4823}.

\bibitem{Kupriyanov:2013jka}
Kupriyanov V.G., Dirac equation on coordinate dependent noncommutative
 space-time, \href{https://doi.org/10.1016/j.physletb.2014.04.006}{\textit{Phys. Lett.~B}} \textbf{732} (2014), 385--390,
 \href{https://arxiv.org/abs/1308.1350}{arXiv:1308.1350}.

\bibitem{Kupriyanov:2008dn}
Kupriyanov V.G., Vassilevich D.V., Star products made (somewhat) easier,
 \href{https://doi.org/10.1140/epjc/s10052-008-0804-2}{\textit{Eur. Phys.~J.~C Part. Fields}} \textbf{58} (2008), 627--637,
 \href{https://arxiv.org/abs/0806.4615}{arXiv:0806.4615}.

\bibitem{Kupriyanov:2015uxa}
Kupriyanov V.G., Vitale P., Noncommutative {${\mathbb R}^d$} via closed star
 product, \href{https://doi.org/10.1007/JHEP08(2015)024}{\textit{J.~High Energy Phys.}} \textbf{2015} (2015), no.~8, 024,
 25~pages, \href{https://arxiv.org/abs/1502.06544}{arXiv:1502.06544}.

\bibitem{PoissonBible}
Laurent-Gengoux C., Pichereau A., Vanhaecke P., Poisson structures,
 \href{https://doi.org/10.1007/978-3-642-31090-4}{\textit{Grundlehren der Mathematischen Wissenschaften}}, Vol.~347, Springer,
 Heidelberg, 2013.

\bibitem{ClassMech}
Lim Y.-K. (Editor), Problems and solutions on mechanics, \textit{Major American
 Universities Ph.D. Qualifying Questions and Solutions}, World Scientific
 Publ., River Edge, NJ, 1994.

\bibitem{Madore:2000en}
Madore J., Schraml S., Schupp P., Wess J., Gauge theory on noncommutative
 spaces, \href{https://doi.org/10.1007/s100520050012}{\textit{Eur. Phys.~J.~C Part. Fields}} \textbf{16} (2000), 161--167,
 \href{https://arxiv.org/abs/hep-th/0001203}{hep-th/0001203}.

\bibitem{Marsden:1999}
Marsden J.E., Ratiu T.S., Introduction to mechanics and symmetry. A~basic
 exposition of classical mecha\-ni\-cal systems, 2nd ed., \href{https://doi.org/10.1007/978-0-387-21792-5}{\textit{Texts in Applied
 Mathematics}}, Vol.~17, Springer-Verlag, New York, 1999.

\bibitem{McCurdy:2009xz}
McCurdy S., Zumino B., Covariant star product for exterior differential forms
 on symplectic manifolds, in Supersymmetry and the Unification of Fundamental
 Interactions, \href{https://doi.org/10.1063/1.3327559}{\textit{AIP Conf. Proc.}}, Vol.~1200, Editors G.~Alverson,
 P.~Nath, B.~Nelson, Amer. Inst. Phys., Melville, NY, 2010, 204--214,
 \href{https://arxiv.org/abs/0910.0459}{arXiv:0910.0459}.

\bibitem{Mezincescu:2000zq}
Mezincescu L., Star operation in quantum mechanics, \href{https://arxiv.org/abs/hep-th/0007046}{hep-th/0007046}.

\bibitem{Micu:2000}
Micu A., Sheikh-Jabbari M.M., Noncommutative {$\Phi^4$} theory at two loops,
 \href{https://doi.org/10.1088/1126-6708/2001/01/025}{\textit{J. High Energy Phys.}} \textbf{2001} (2001), no.~1, 025, 45~pages,
 \href{https://arxiv.org/abs/hep-th/0008057}{hep-th/0008057}.

\bibitem{Moscovici}
Moscovici H., Local index formula and twisted spectral triples, in Quanta of
 Maths, \textit{Clay Math. Proc.}, Vol.~11, Amer. Math. Soc., Providence, RI,
 2010, 465--500, \href{https://arxiv.org/abs/0902.0835}{arXiv:0902.0835}.

\bibitem{Moyal:1949}
Moyal J.E., Quantum mechanics as a statistical theory, \href{https://doi.org/10.1017/S0305004100000487}{\textit{Proc. Cambridge
 Philos. Soc.}} \textbf{45} (1949), 99--124.

\bibitem{Mueller-Hoissen}
M\"{u}ller-Hoissen F., Noncommutative geometries and gravity, in Recent
 Developments in Gravitation and Cosmology, \href{https://doi.org/10.1063/1.2902778}{\textit{AIP Conf. Proc.}}, Vol.~977, Amer. Inst. Phys., Melville, NY, 2008, 12--29, \href{https://arxiv.org/abs/0710.4418}{arXiv:0710.4418}.

\bibitem{Ovsienko}
Ovsienko V., Tabachnikov S., Projective differential geometry old and new. From
 the Schwarzian derivative to the cohomology of diffeomorphism groups,
 \href{https://doi.org/10.1063/1.2902778}{\textit{Cambridge Tracts in Mathematics}}, Vol.~165, Cambridge University
 Press, Cambridge, 2005.

\bibitem{Pengpan:2000kd}
Pengpan T., Xiong X., Remarks on the noncommutative {W}ess--{Z}umino model,
 \href{https://doi.org/10.1103/PhysRevD.63.085012}{\textit{Phys. Rev.~D}} \textbf{63} (2001), 085012, 7~pages,
 \href{https://arxiv.org/abs/hep-th/0009070}{hep-th/0009070}.

\bibitem{Poulain:2018mcm}
Poulain T., Wallet J.-C., $\kappa$-Poincar{\'{e}} invariant quantum field
 theories with {K}ubo--{M}artin--{S}chwinger weight, \href{https://doi.org/10.1103/PhysRevD.98.025002}{\textit{Phys. Rev.~D}}
 \textbf{98} (2018), 025002, 22~pages, \href{https://arxiv.org/abs/1801.02715}{arXiv:1801.02715}.

\bibitem{Reshetikhin}
Reshetikhin N., Multiparameter quantum groups and twisted quasitriangular
 {H}opf algebras, \href{https://doi.org/10.1007/BF00626530}{\textit{Lett. Math. Phys.}} \textbf{20} (1990), 331--335.

\bibitem{Rivasseau:2007a}
Rivasseau V., Non-commutative renormalization, in Quantum Spaces~--
 Poincar\'{e} Seminar 2007, \href{https://doi.org/10.1007/978-3-7643-8522-4_2}{\textit{Progress in Mathematical Physics}},
 Vol.~53, Editors B.~Duplantier, V.~Rivasseau, Birkh\"{a}user Verlag, Basel,
 2008, 19--107, \href{https://arxiv.org/abs/0705.0705}{arXiv:0705.0705}.

\bibitem{ClaudeRoger}
Roger C., Gerstenhaber and {B}atalin--{V}ilkovisky algebras; algebraic,
 geometric, and physical aspects, \textit{Arch. Math. (Brno)} \textbf{45}
 (2009), 301--324.

\bibitem{Rosa:2012pr}
Rosa L., Vitale P., On the {$\star$}-product quantization and the {D}uflo map
 in three dimensions, \href{https://doi.org/10.1142/S0217732312502070}{\textit{Modern Phys. Lett.~A}} \textbf{27} (2012),
 1250207, 15~pages, \href{https://arxiv.org/abs/1209.2941}{arXiv:1209.2941}.

\bibitem{Schenkel:2012zu}
Schenkel A., Noncommutative gravity and quantum field theory on noncommutative
 curved spacetimes, Ph.D.~Thesis, W{\"u}rzburg University, 2011,
 \href{https://arxiv.org/abs/1210.1115}{arXiv:1210.1115}.

\bibitem{Schenkel:2010sc}
Schenkel A., Uhlemann C.F., Field theory on curved noncommutative spacetimes,
 \href{https://doi.org/10.3842/SIGMA.2010.061}{\textit{SIGMA}} \textbf{6} (2010), 061, 19~pages, \href{https://arxiv.org/abs/1003.3190}{arXiv:1003.3190}.

\bibitem{Schweber}
Schweber S.S., An introduction to relativistic quantum field theory, \textit{Dover
 Books on Physics}, Dover Publications, New York, 2005.

\bibitem{Steinacker:2010rh}
Steinacker H., Emergent geometry and gravity from matrix models: an
 introduction, \href{https://doi.org/10.1088/0264-9381/27/13/133001}{\textit{Classical Quantum Gravity}} \textbf{27} (2010), 133001,
 46~pages, \href{https://arxiv.org/abs/1003.4134}{arXiv:1003.4134}.

\bibitem{Steinacker:2012ct}
Steinacker H., The curvature of branes, currents and gravity in matrix models,
 \href{https://doi.org/10.1007/JHEP01(2013)112}{\textit{J.~High Energy Phys.}} \textbf{2013} (2013), no.~1, 112, 28~pages,
 \href{https://arxiv.org/abs/1210.8364}{arXiv:1210.8364}.

\bibitem{Stephani:2003tm}
Stephani H., Kramer D., MacCallum M., Hoenselaers C., Herlt E., Exact solutions
 of {E}instein's field equations, 2nd~ed., \href{https://doi.org/10.1017/CBO9780511535185}{\textit{Cambridge Monographs on
 Mathematical Physics}}, Cambridge University Press, Cambridge, 2003.

\bibitem{straumann}
Straumann N., General relativity, 2nd ed., \href{https://doi.org/10.1007/978-94-007-5410-2}{\textit{Graduate Texts in Physics}}, Springer,
 Dordrecht, 2013.

\bibitem{Struckmeier:2005}
Struckmeier J., Hamiltonian dynamics on the symplectic extended phase space for
 autonomous and non-autonomous systems, \href{https://doi.org/10.1088/0305-4470/38/6/006}{\textit{J.~Phys.~A: Math. Gen.}}
 \textbf{38} (2005), 1257--1278.

\bibitem{Jambor:2004kc}
Sykora A., Jambor C., Realization of algebras with the help of
 $\star$-products, \href{https://arxiv.org/abs/hep-th/0405268}{hep-th/0405268}.

\bibitem{Szabo:2001}
Szabo R.J., Quantum field theory on noncommutative spaces, \href{https://doi.org/10.1016/S0370-1573(03)00059-0}{\textit{Phys. Rep.}}
 \textbf{378} (2003), 207--299, \mbox{\href{https://arxiv.org/abs/hep-th/0109162}{hep-th/0109162}}.

\bibitem{Szabo:2006wx}
Szabo R.J., Symmetry, gravity and noncommutativity, \href{https://doi.org/10.1088/0264-9381/23/22/R01}{\textit{Classical Quantum
 Gravity}} \textbf{23} (2006), R199--R242, \href{https://arxiv.org/abs/hep-th/0606233}{hep-th/0606233}.

\bibitem{Vassilevich:2009cb}
Vassilevich D.V., Diffeomorphism covariant star products and noncommutative
 gravity, \href{https://doi.org/10.1088/0264-9381/26/14/145010}{\textit{Classical Quantum Gravity}} \textbf{26} (2009), 145010,
 8~pages, \href{https://arxiv.org/abs/0904.3079}{arXiv:0904.3079}.

\bibitem{Vassilevich:2010he}
Vassilevich D.V., Tensor calculus on noncommutative spaces, \href{https://doi.org/10.1088/0264-9381/27/9/095020}{\textit{Classical
 Quantum Gravity}} \textbf{27} (2010), 095020, 16~pages, \href{https://arxiv.org/abs/1001.0766}{arXiv:1001.0766}.

\bibitem{Waldmann:2007}
Waldmann S., Poisson-Geometrie und Deformationsquantisierung. Eine
 Einf\"uhrung, \href{https://doi.org/10.1007/978-3-540-72518-3}{Springer-Verlag}, Berlin~-- Heidelberg, 2007.

\bibitem{Wallet:2007em}
Wallet J.-C., Noncommutative induced gauge theories on {M}oyal spaces,
 \href{https://doi.org/10.1088/1742-6596/103/1/012007}{\textit{J.~Phys. Conf. Ser.}} \textbf{103} (2008), 012007, 20~pages,
 \href{https://arxiv.org/abs/0708.2471}{arXiv:0708.2471}.

\bibitem{Witten:1990wb}
Witten E., A note on the antibracket formalism, \href{https://doi.org/10.1142/S0217732390000561}{\textit{Modern Phys. Lett.~A}}
 \textbf{5} (1990), 487--494.

\bibitem{Wohlgenannt:2006dx}
Wohlgenannt M., Non-commutative geometry and physics, \textit{Ukr.~J. Phys.}
 \textbf{55} (2010), 5--14, \href{https://arxiv.org/abs/hep-th/0602105}{hep-th/0602105}.

\bibitem{Wulkenhaar:2006si}
Wulkenhaar R., Field theories on deformed spaces, \href{https://doi.org/10.1016/j.geomphys.2005.04.019}{\textit{J.~Geom. Phys.}}
 \textbf{56} (2006), 108--141.

\end{thebibliography}
\end{document}